\documentclass[twocolumn,preprintnumbers,amsmath,amssymb,aps,prb,longbibliography]{revtex4-1}
\usepackage{graphicx}
\begin{document}

\title{Shear Banding, Intermittency, Jamming  and Dynamic Phases For Skyrmions in Inhomogeneous Pinning Arrays}
 
\author{C. Reichhardt and C. J. O. Reichhardt}
\affiliation{Theoretical Division and Center for Nonlinear Studies,
Los Alamos National Laboratory, Los Alamos, New Mexico 87545, USA}

\date{\today}

\begin{abstract}
  We examine driven skyrmion dynamics in systems with inhomogeneous pinning where a strip of strong pinning coexists with a region containing no pinning.  For driving parallel to the strip, we find that the initial skyrmion motion is confined to the unpinned region and the skyrmion Hall angle is zero.  At larger drives, a transition occurs to a phase in which motion also appears within the pinned region, creating a shear band in the skyrmion velocity, while the skyrmion Hall angle is still zero.  As the drive increases further, the flow becomes disordered and the skyrmion Hall angle increases with drive until saturating at the highest drives when the system transitions into a moving crystal phase.   The different dynamic phases are associated with velocity and density gradients across the pinning boundaries.  We map out the dynamic phases as a function of pinning strength, skyrmion density, and Magnus force strength, and correlate the phase boundaries with features in the velocity-force curves and changes in the local and global ordering of the skyrmion structure.  For large Magnus forces, the shear banding instability is replaced by large scale intermittent flow in the pinned region accompanied by simultaneous motion perpendicular to the direction of the drive, which appears as oscillations in the transport curves.  We also examine the case of a drive applied perpendicular to the strip, where we find a jamming effect in which the skyrmion flow is blocked by skyrmion-skyrmion interactions until the drive is large enough to induce plastic flow.
\end{abstract}

\maketitle

\vskip 2pc

\section{Introduction}
Skyrmions in magnetic systems were discovered in MnSi in 2009 \cite{Muhlbauer09}, 
and since that time skyrmions have been found in an increasing variety of systems,
including materials in which the skyrmions are stable at room temperature
\cite{Yu10,Nagaosa13,Woo16,Soumyanarayanan17,Fert17}. 
In samples with weak quenched disorder,
the skyrmions form a triangular lattice
and can be set into motion with an applied current
\cite{Schulz12,Yu12,Iwasaki13,Lin13a,Okuyama19}.
When
disorder is present,
there is a finite depinning threshold for skyrmion motion
and there can be different types 
of flow such as plastic or ordered 
as well as transitions between different types
of moving phases
\cite{Nagaosa13,Woo16,Fert17,Schulz12,Iwasaki13,Okuyama19,Liang15,Reichhardt15,Jiang17,Legrand17,Sato19}.
The onset of motion and the dynamic phase transitions
are correlated with changes in the
skyrmion velocity-force curves,
the skyrmion flow patterns,
the structure factor, and the velocity noise spectra \cite{Diaz17}.   

Although skyrmions have many similarities to other systems that are known to exhibit 
depinning,
such as vortices in type-II superconductors or
colloidal particles on rough landscapes \cite{Reichhardt17}, 
they also have
the unique feature of
a strong non-dissipative Magnus force which creates
velocity components that are perpendicular to the
forces produced by the drive, pinning, and interaction with other skyrmions.
The ratio of the Magnus force to the dissipation
can range from a few percent to
up to a factor of 10 or more \cite{Nagaosa13,EverschorSitte18,Iwasaki13}.
One consequence of the Magnus force
is that under a drive,
skyrmions display a strong gyroscopic motion that produces
a finite Hall angle
known
as the skyrmion Hall angle
\cite{Nagaosa13,Reichhardt15,Jiang17,Legrand17,Zang11,EverschorSitte14,Kim19}.
The gyroscopic 
motion
generates spiraling skyrmion orbits in confining potentials or pinning sites
\cite{Liu13,Muller15,Buttner15} which have been proposed
to be one reason why the pinning of
skyrmions is often weak,
since a skyrmion can spiral around pinning
sites rather than becoming trapped \cite{Nagaosa13,Schulz12,Iwasaki13}.
There are, however, other cases in which
the effect of pinning on skyrmions can be strong 
\cite{Woo16,Jiang17,Legrand17}.

The skyrmion
Hall angle is constant in the absence of pinning or disorder,
but in the presence of disorder it develops
a dependence on drive or velocity,
starting at a value of zero just at the depinning threshold and
gradually increasing with increasing skyrmion velocity
until saturating at
high drives to a value close to that found in the clean limit 
\cite{Reichhardt15,Jiang17,Legrand17,Reichhardt16,Kim17,Litzius17,Woo18,Juge19,Zeissler19}. 
Particle based  \cite{Reichhardt15,Muller15,Reichhardt16,Reichhardt19,Brown19}
and continuum based \cite{Legrand17,Muller15,Kim17,Juge19}  simulations
show that the drive dependence of 
the skyrmion Hall angle
is a result of the skyrmion-pin interactions.
In regimes of collective skyrmion motion in
the presence of quenched disorder, the flow above depinning can be
elastic when the pinning is weak, with all skyrmions maintaining the
same neighbors as they move
\cite{Yu12,Iwasaki13,Reichhardt15,Reichhardt19},
or it can be plastic \cite{Reichhardt15,Legrand17,Reichhardt16,Koshibae18,Montoya18},
with a combination of pinned and moving skyrmions.
In some cases, the skyrmions
exhibit intermittent or avalanche-like flow, in which
sudden bursts of motion
are interspersed among intervals in which no motion occurs \cite{Diaz18,Singh19}.   

In systems that exhibit depinning,
such as superconducting vortices \cite{Bhattacharya93,Olson98a},
colloidal assemblies \cite{Bohlein12,Tierno12a},
electron crystals \cite{Reichhardt01,Brussarski18},
or charge density waves \cite{Gruner88}, the disorder 
is often homogeneous on long length scales,
so that on average, the same depinning 
threshold occurs throughout the sample and there is
a single well defined depinning drive.
It is also possible for samples to contain
strongly inhomogeneous pinning,
where strong pinning in some portions of the sample coexists
with other regions
in which the pinning is absent or weak.
This type of pinning can arise naturally
if the system
has large scale inhomogeneities,
or it can be created artificially
using lithographic techniques by writing a pinning array into
only selected
portions of the sample
while other portions of the sample remain pin-free.
In colloidal systems
with inhomogeneous pinning, it was shown that
flow occurs in the unpinned regions and that
shear banding effects arise 
when a portion of the colloids in the unpinned region
are either pinned or moving more slowly due to interactions with
neighboring pinned colloids,
creating a velocity gradient \cite{Seshadri93,Seshadri93a}.
Other studies of colloidal systems containing a strip of pinned
colloids revealed that the system effectively freezes
from the pinned strip outward into the bulk \cite{Nagamanasa16}.
In superconducting systems, coexisting regions
of strong pinning and weak pinning were created
by patterned irradiation in order to study
shearing effects in the vortex lattice when flow initiates
in the unpinned regions and strong velocity gradients appear
\cite{Marchetti90,Marchetti99,Basset01,Kwok02,Banerjee03}.
It is also possible to create spatially inhomogeneous pinning
by creating
large scale thickness modulations
\cite{Besseling05,Togawa05,Yu07,Dobrovolskiy12},
diluting periodic pinning arrays  \cite{Reichhardt07a,Kemmler09},
or by selecting a sample with strong edge pinning but weak bulk pinning \cite{Kimmel19}.
Another way to
introduce inhomogeneous depinning thresholds
is by creating
a  gradient in the number or size of the pinning sites;
in this case, the dynamics depend on whether the system is driven
parallel or perpendicular to the gradient
\cite{Ray14b,Wang13,Guenon13,Motta13,Reichhardt19a}.
Other studies
showed that inhomogeneous pinning can lead to a number of interesting dynamical 
effects such as negative mobility \cite{Xu07} and ratchet motion
\cite{Gillijns07,Wu10}. 
In charge density wave systems with inhomogeneous pinning,
depinning first occurs in the weak pinning region,
creating a shearing effect in the more strongly pinned region \cite{Isakovic06}. 
In systems with
spatially inhomogeneous pinning,
application of a drive tends to
create velocity gradients which lead to 
the formation of dislocations,
the emergence of liquid phases,
or the coexistence of liquid and solid phases.
Many of these phases are similar to those
found in systems with homogeneous pinning or no pinning
when the driving is inhomogeneous,
such as in Corbino geometries for superconducting vortices, where
different
phases appear such as a solid flow
in which
the vortex lattice rotates as a rigid body,
as well as a shear banded state
at higher drives
\cite{Lopez99,Miguel03,Furukawa06,Lin10,Miguel11,Rosen13,Kawamura15}.

In the studies performed with inhomogeneous pinning up until now,
the dynamics has been exclusively overdamped,
so it is not known when happens when there is a finite Magnus force.
Since the Magnus force mixes the velocity components from external drives,
one would expect rather different results to appear compared to what
is found in the overdamped systems.
Another interesting effect is that shear banding
could arise for driving either parallel or perpendicular to the inhomogeneous pinning
regions due to the Magnus force.  
There have been some studies of skyrmions under inhomogeneous drives
in the absence of pinning which produced evidence
for rigid flow, disordered flow, and shear banding effects
\cite{Mochizuki14,Pollath17,Zhang18}.

Here we examine skyrmion dynamics in a system
where the external drive is uniform but
the pinning is inhomogeneous, with a region of strong pinning in the form of
a strip coexisting 
with a region where there is no pinning. 
When the drive is parallel to the strip, we find that 
the skyrmions first move in the unpinned region
and that the skyrmion Hall effect is suppressed.
As the drive is increased,
flow occurs in both regions
with a velocity gradient in the pinned region,
but the skyrmion Hall angle is still zero. 
At higher drives, a shear-induced disordered or liquid phase appears
due to a proliferation of topological defects in the skyrmion lattice,
and the skyrmion Hall angle becomes finite,
while at very high drives, the flow becomes uniform, the skyrmions
form a moving liquid or moving crystal state, and the skyrmion Hall angle
reaches a saturation value.
For a drive applied perpendicular to the strip,
when the Magnus force is large we find that skyrmions from the unpinned region
enter the pinned region in 
avalanches,
creating a density gradient analogous to the
Bean state found in type-II superconductors \cite{Bean64,Reichhardt95}.
Here the skyrmions accumulate along the edge of the pinned region
and form a jammed state in which
the repulsive interactions from 
the skyrmions in the pinned region block the flow of the skyrmions
in the unpinned region.
In previous work on
skyrmion motion in inhomogeneous pinning,
we considered only the case of skyrmion flow in the unpinned region \cite{Reichhardt19c}.
Here we 
expand on this to study the entire range
of dynamics and the interplay of motion in both the unpinned and pinned regions as
well as driving in different directions.
Although we focus on skyrmions, our results should be applicable to 
other types of systems with a Magnus force or gyroscopic coupling in
the presence of inhomogeneous disorder. Examples of such 
systems include colloidal rotators \cite{Han17}, 
magnetically driven colloids \cite{Loehr18,Denisov18,MassanaCid19},
vortices in superfluids \cite{Wlazlowski16,Griffin19},
and chiral active matter states \cite{vanZuiden16,Reichhardt19d,Reichhardt19e}.

\section{Simulation}

\begin{figure}
\includegraphics[width=3.5in]{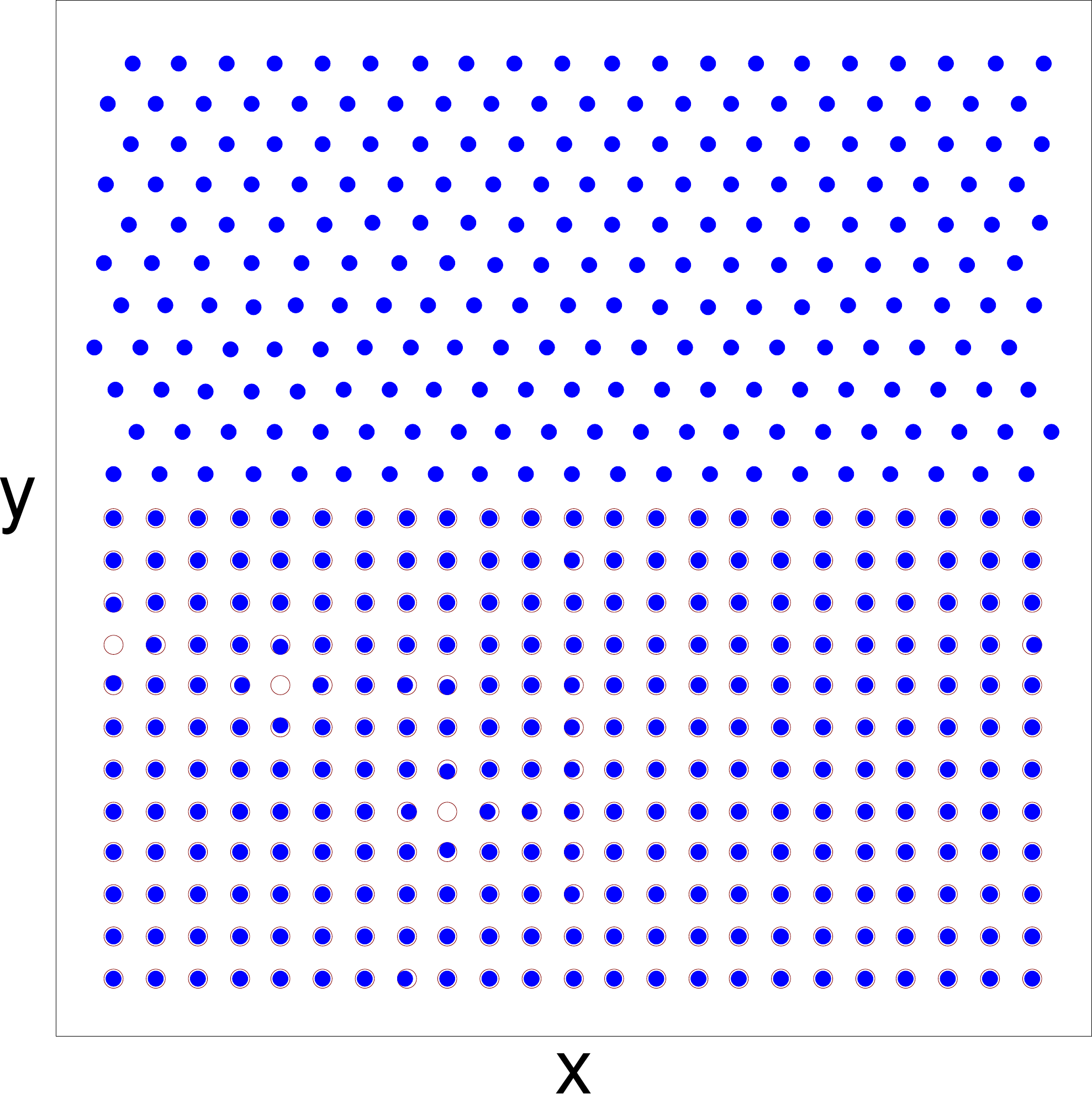}
\caption{Image of the system showing skyrmion positions (blue dots)
  and pinning site locations (open red circles).  There are half as many skyrmions
  as pinning sites and only the lower half of the sample contains pinning.
  The drive can be 
applied along either the $x$ or the $y$ direction.} 
\label{fig:1}
\end{figure}

We 
consider a two-dimensional system of size $L \times L$ with periodic boundary 
conditions in the $x$ and $y$-directions
where the skyrmions are modeled as particles with
skyrmion-skyrmion and skyrmion-pinning interactions 
based a modified Theile equation
\cite{Reichhardt15,Reichhardt16,Lin13,Brown19a,Xiong19}.  
Half of the sample is pin-free, and the other half of the sample
contains a square array of pinning sites, as illustrated in Fig.~\ref{fig:1}.
The pinning sites
are modeled as
finite range parabolic traps with lattice constant $a$, pinning radius $r_{p}$, and 
maximum strength $F_{p}$.
The sample contains $N_{p}$ pinning sites and $N$ skyrmions, and we
focus on the case of $N/N_{p} = 2.0$,
so that under equilibrium conditions the skyrmion density is uniform,
with half of the skyrmions in the pinned region and half of the skyrmions in the
unpinned region.
We define the matching density $n_\phi=2N_p/L^2$ to be the density at which
the number of skyrmions would match the number of pinning sites if the entire
sample were filled with the same density of pinning as the pinned region.
Throughout this work we consider $n_\phi=0.4$.

The dynamics of skyrmion $i$ is governed  by the following equation of motion:      
\begin{equation} 
\alpha_d {\bf v}_{i} + \alpha_m {\hat z} \times {\bf v}_{i} =
{\bf F}^{ss}_{i} + {\bf F}^{D}_{i} .
\end{equation}
Here the repulsive skyrmion-skyrmion force is  
${\bf F}_{i} = \sum^{N}_{j=1}K_{1}(r_{ij}){\hat {\bf r}_{ij}}$,
where $r_{ij} = |{\bf r}_{i} - {\bf r}_{j}|$, $\hat {\bf r}_{ij}=({\bf r}_i-{\bf r}_j)/r_{ij}$, and
the modified Bessel function $K_{1}(r)$ falls off exponentially for large $r$.
A uniform driving force ${\bf F}^{D}=F_D{\hat \alpha}$ is applied to all skyrmions
in either the $x$-direction ($\alpha=x$), parallel to the pinning strip,
or in the $y$-direction ($\alpha=y$), perpendicular to the pinning strip.
The skyrmion velocity is ${\bf v}$, and 
the damping term $\alpha_d$  aligns the velocity
in the direction of the net applied forces.
The Magnus term, with coefficient $\alpha_{m}$,
creates velocities that are perpendicular to the net external forces.
When the Magnus term is finite, in the 
absence 
of pinning the skyrmions  move at an angle with respect to the driving
force
given by the intrinsic skyrmion Hall angle
$\theta^{\rm int}_{sk} = \tan^{-1}(\alpha_{m}/\alpha_{d})$.

The initial skyrmion positions are obtained through simulated annealing,
after which we apply a drive
which we increase in increments of $\delta F_D$
with a fixed number of simulation time steps spent at each value of $F_D$.
For each value of the drive,
we measure the average velocity both parallel,
$\langle V_{||}\rangle=N^{-1}\sum_i^N {\bf v}_i \cdot {\bf \hat x}$,
and perpendicular,
$\langle V_\perp\rangle=N^{-1}\sum_i^N {\bf v}_i \cdot {\bf \hat y}$,
to the pinning strip.
The measured skyrmion Hall angle
is $\theta_{sk} = \tan^{-1}(\langle V_\perp\rangle/\langle V_{||}\rangle)$.
For the studies reported here
we typically use increments of
$\delta F_D=0.00025$ and we wait $2000$ simulation time steps between force
increments.
The velocity-force characteristics and average measured quantities
do not change for smaller force increments or longer waiting times.     

\section{Shearing Dynamics for Parallel Driving}

\begin{figure}
\includegraphics[width=3.5in]{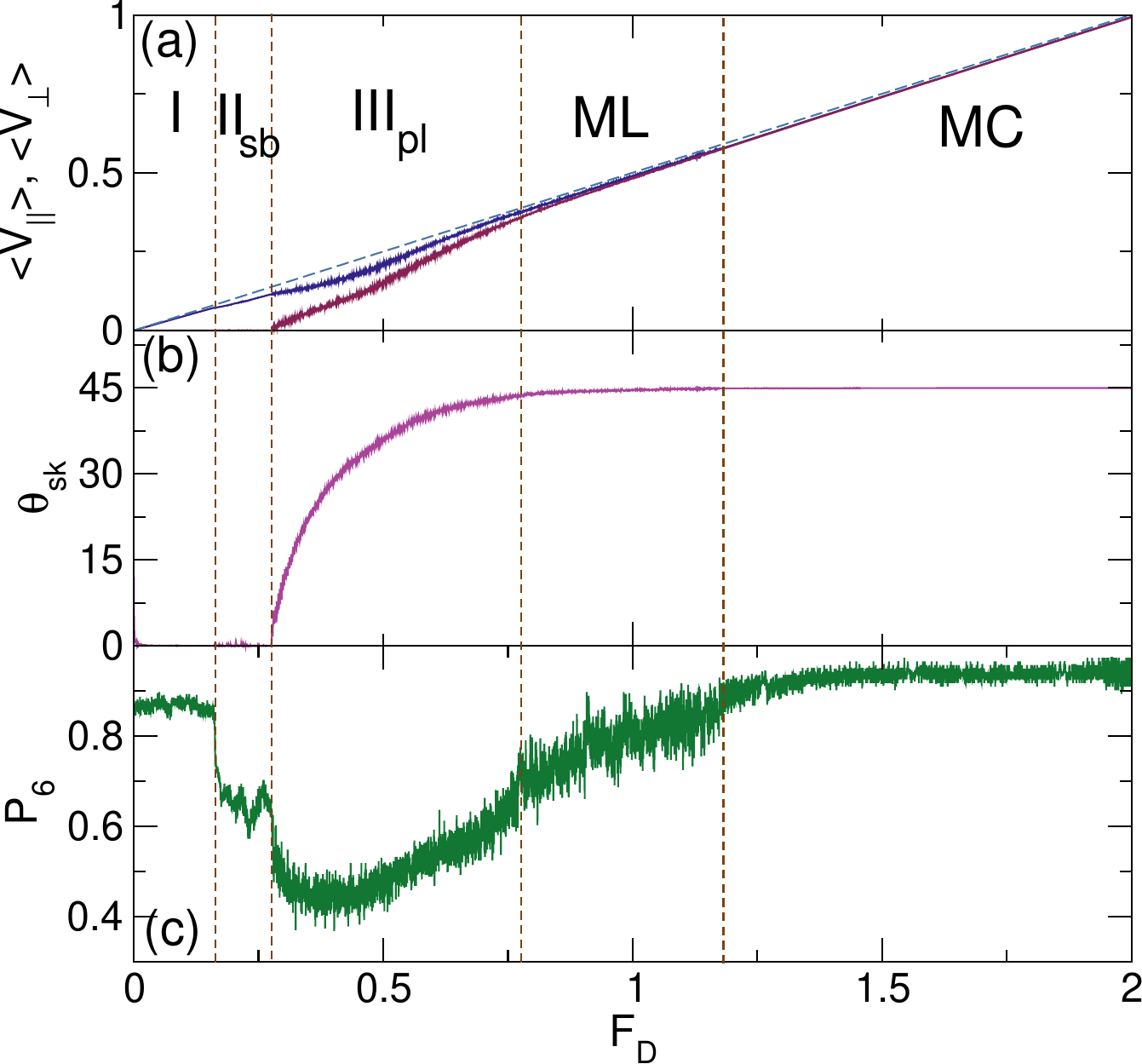}
\caption{(a)
  $\langle V_{||}\rangle$ (blue solid line) and
  $\langle V_\perp\rangle$ (red line) vs $F_{D}$ for a sample with
  $x$ direction driving at
  $F_{p} = 0.75$ and $\alpha_{m}/\alpha_{d} = 1.0$.
  Vertical dashed lines indicate the four dynamical phases:
  I, longitudinal flow in the $x$ direction in only the pin-free channel;
  II$_{sb}$, the shear banding phase;
  III$_{pl}$, plastic flow;
  ML, a moving liquid;
  and MC, a moving crystal.
  (b) The corresponding skyrmion Hall angle $\theta_{sk}$ vs $F_{D}$
  showing that $\theta_{sk}$ increases from zero in
  phase III$_{pl}$ and reaches a saturation value in the ML and MC phases.      
  (c) The corresponding fraction of six-fold coordinated skyrmions
  $P_6$ vs $F_{D}$, showing changes
across each of the four dynamic phase transitions.}   
\label{fig:2}
\end{figure}

\begin{figure}
\includegraphics[width=3.5in]{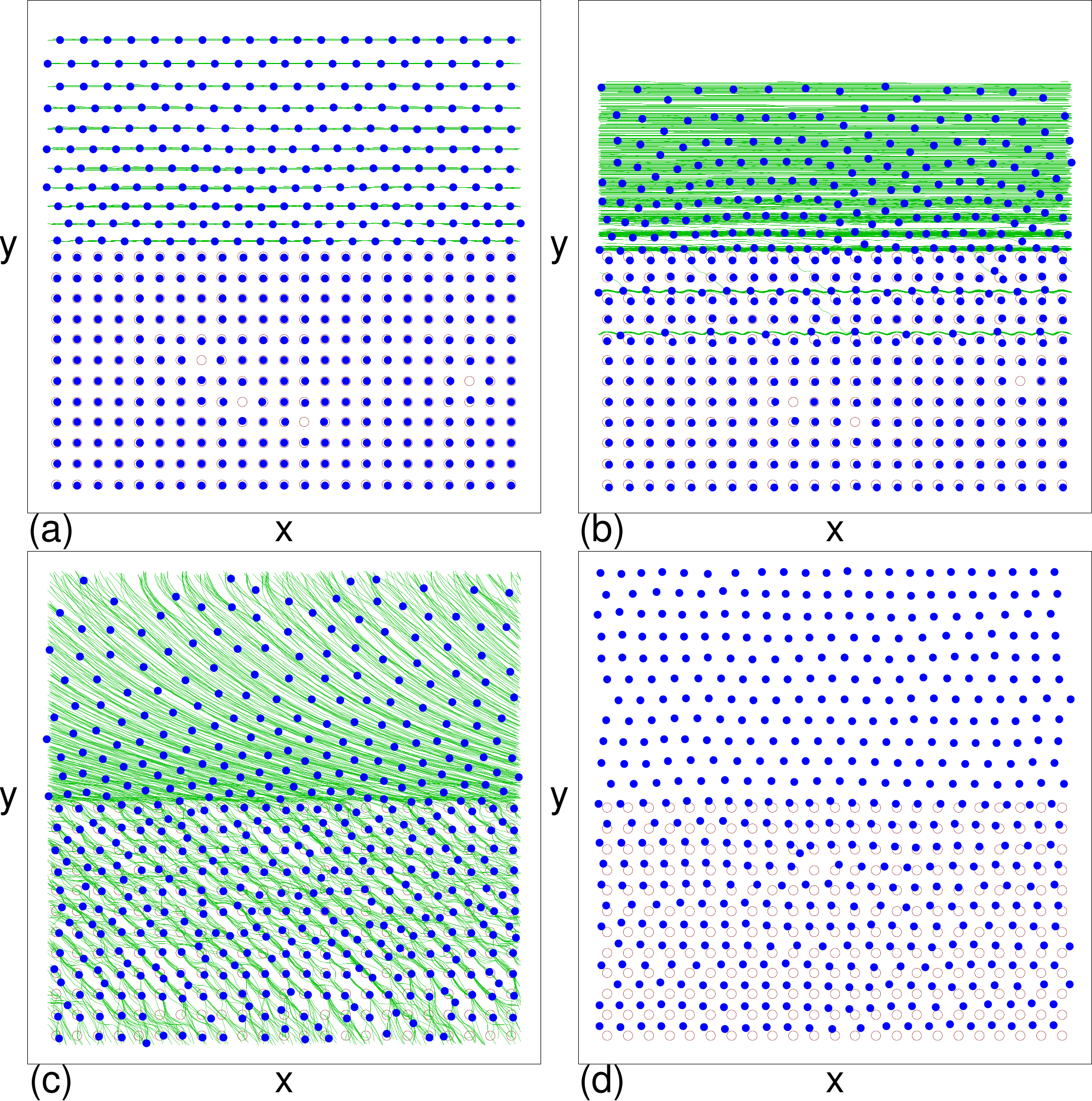}
\caption{
  Skyrmion locations (blue dots),  pinning site locations (open circles), and
  skyrmion trajectories (green lines) during a fixed time interval for the system
  in Fig.~\ref{fig:2} with $x$ direction driving at $F_p=0.75$
  and $\alpha_m/\alpha_d=1.0$.
  (a) Phase I at $F_D=0.1$, where the flow is only in the unpinned region. 
  (b) Phase II$_{sb}$ at $F_D=0.2$, where there is flow in both the unpinned
  and pinned regions but the skyrmion Hall angle is zero.
  (c) Phase III$_{pl}$ at $F_D=0.5$ where the skyrmion Hall angle is finite.
  (d) The pinning sites and skyrmion locations
  without trajectories in the moving crystal phase MC at $F_D=1.5$.}  
\label{fig:3}
\end{figure}

\begin{figure}
\includegraphics[width=3.5in]{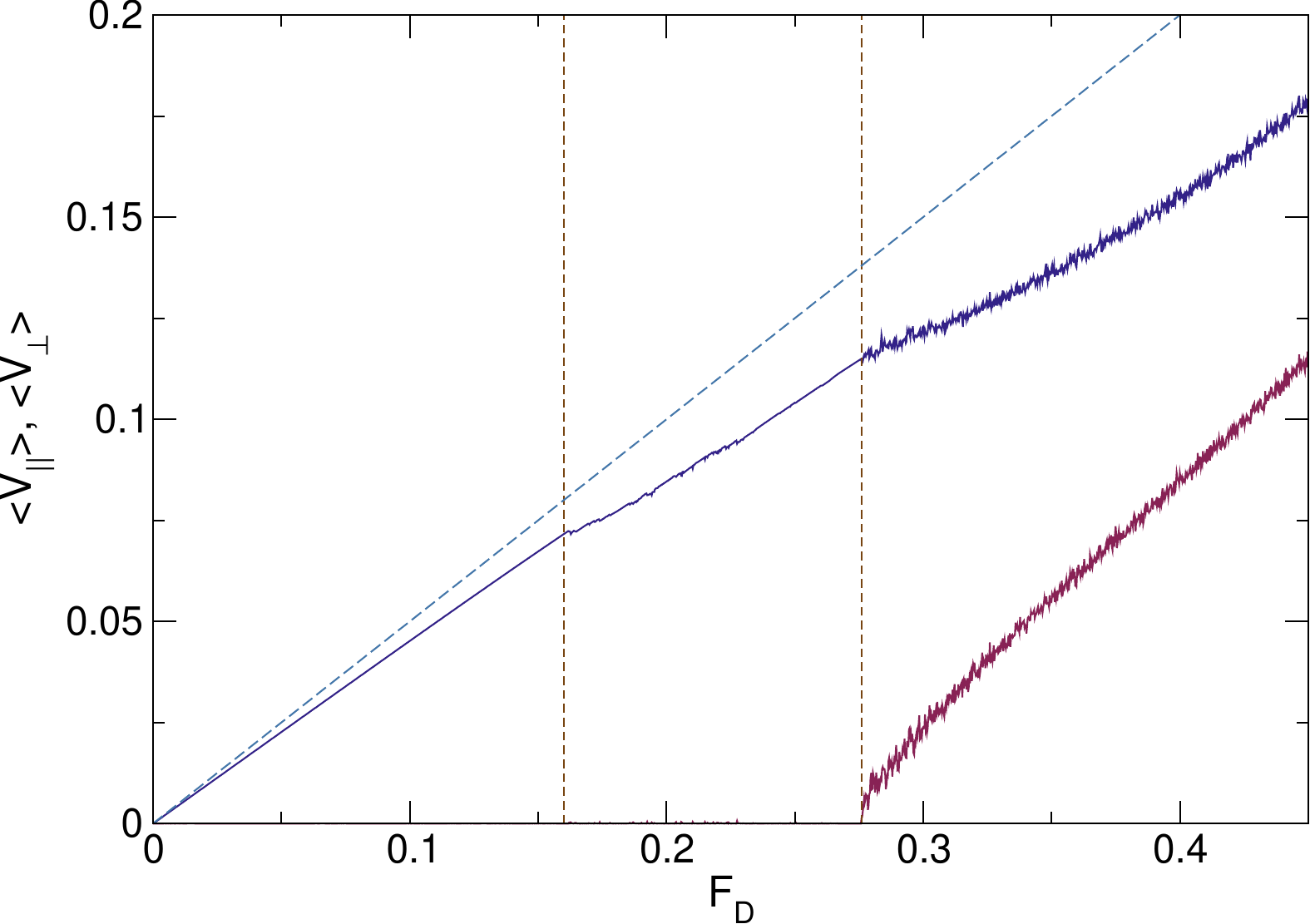}
\caption{ A blowup of the $\langle V_{||}\rangle$ (blue solid line)
  and $\langle V_{\perp}\rangle$ (red line)
vs $F_{D}$ curves from Fig.~\ref{fig:2}(a) highlighting the I-II$_{sb}$ transition
at $F_D=0.16$.
The blue dashed line is the velocity $\langle V_{||}\rangle$ in the pin-free limit.}  
\label{fig:4}
\end{figure}

We first consider the case where the skyrmions are driven in the $x$ direction,
parallel to the pinning stripe.
In Fig.~\ref{fig:2}(a) we plot
$\langle V_{||}\rangle$ and $\langle V_\perp\rangle$ versus
$F_{D}$ for a sample with
$F_{p} = 0.75$
and $\alpha_{m}/\alpha_{d} = 1.0$.
In the absence of pinning,
the skyrmions form a triangular lattice that moves at
an angle of $45^\circ$ with respect to the $x$-axis, as indicated by the
dashed blue line.
Figure~\ref{fig:2}(b) shows the corresponding
measured skyrmion Hall angle $\theta_{sk}$ versus $F_{D}$.
As indicated in Fig.~\ref{fig:2}(a), we identify five dynamic phases. 
In phase I, $\langle V_{||}\rangle$ is finite and $\langle V_\perp\rangle = 0.0$,
and the skyrmions flow
only in the unpinned portions of the sample in the direction of the drive.
This motion is illustrated
in Fig.~\ref{fig:3}(a),
where the skyrmions flow elastically inside the pin-free region
and the skyrmion Hall angle is zero.   
As the drive increases,
the system enters the shear banding phase
II$_{sb}$
where the flow is still only along the $x$-direction but skyrmions move both
in the unpinned region
and in the pinned region,
as shown in Fig.~\ref{fig:3}(b). 
In phase II$_{sb}$, the skyrmions in the unpinned region begin
to accumulate along one edge of the pinned region
due to the Magnus force, which acts along the $y$ direction perpendicular
to the pin-free region.
For $F_{D} > 0.28$,
in phase III$_{pl}$ or the plastic flow state
there is now flow in both the $x$ and $y$ directions and
skyrmions can move across the entire 
pinned strip.
Within the pinned region there
is a combination of pinned and moving skyrmions which creates
the disordered motion illustrated in Fig.~\ref{fig:3}(c).
The skyrmion Hall angle $\theta_{sk}$ increases from zero
in phase III$_{pl}$
and it begins to saturate once $F_{D}/F_{p} \gtrsim 1.0$.
When $0.75 < F_{D} < 1.18$,
all the skyrmions are moving since $F_{D} > F_{p}$;
however, the flow is still disordered,
and the system is in the moving liquid phase ML.
For $F_{D} > 1.18$, there is a transition to a moving 
crystal (MC) phase of the type shown in Fig.~\ref{fig:3}(d).
Within the ML phase, the increase of $\theta_{sk}$ with
increasing $F_{D}$ is less rapid compared to phase III$_{pl}$,
while within the MC phase,
$\theta_{sk}$ is constant at the clean limit value
of $\theta_{sk}\approx 45^\circ$,
and fluctuations in
$\langle V_{||}\rangle$ and $\langle V_{\perp}\rangle$ are
strongly reduced.
The vertical lines in Fig.~\ref{fig:2} indicate the transitions between
the five different phases.
To better highlight the
I-II$_{sb}$ transition, in Fig.~\ref{fig:4} we
show a blowup of
$\langle V_{||}\rangle$  and $\langle V_{\perp}\rangle$ versus
$F_{D}$,
where the blue dashed line indicates the expected value of
$\langle V_{||}\rangle$ in a pin-free system.
Across the I-II$_{sb}$ transition, there is a change in the slope 
of $\langle V_{||}\rangle$ as a function of $F_{D}$ 
when skyrmions in the pin free region start to
accumulate along the edge of the pinned region
and the skyrmions in the pinned region begin to move in the $x$-direction.        

Another method for characterizing the different phases is to measure the
fraction $P_6$ of six-fold coordinated skyrmions.
We generate a Voronoi construction from the skyrmion positions to identify
the coordination number $z_i$ of each skyrmion, and then obtain
$P_6=N^{-1}\sum_{i}^N \delta (6-z_i)$.
In Fig.~\ref{fig:2}(c),
$P_{6}$ versus $F_{D}$
has a signature at all four phase transitions.
In phase I, the ordering is mostly triangular due to the arrangement of the
skyrmions in the pin-free region, as shown in Fig.~\ref{fig:1}, and $P_6$ also picks
up some finite weight in the pinned region due to small distortions that can
produce short additional sides in the Voronoi polygons.
When the system enters the II$_{sb}$ phase, there is a drop in 
$P_{6}$ due to the formation of dislocations that glide along the $x$-direction.
There is
another sharp drop
in $P_{6}$ at the onset of the
strongly disordered III$_{pl}$,
where topological defects proliferate.
In the ML phase, there is more topological order
and $P_{6}$ increases to $P_6 \approx 0.825$,
but a number of dislocations are still present in the sample.
Upon entering the MC phase,
$P_{6} \approx 0.95$ since the skyrmions exhibit
crystalline order.

\begin{figure}
\includegraphics[width=3.5in]{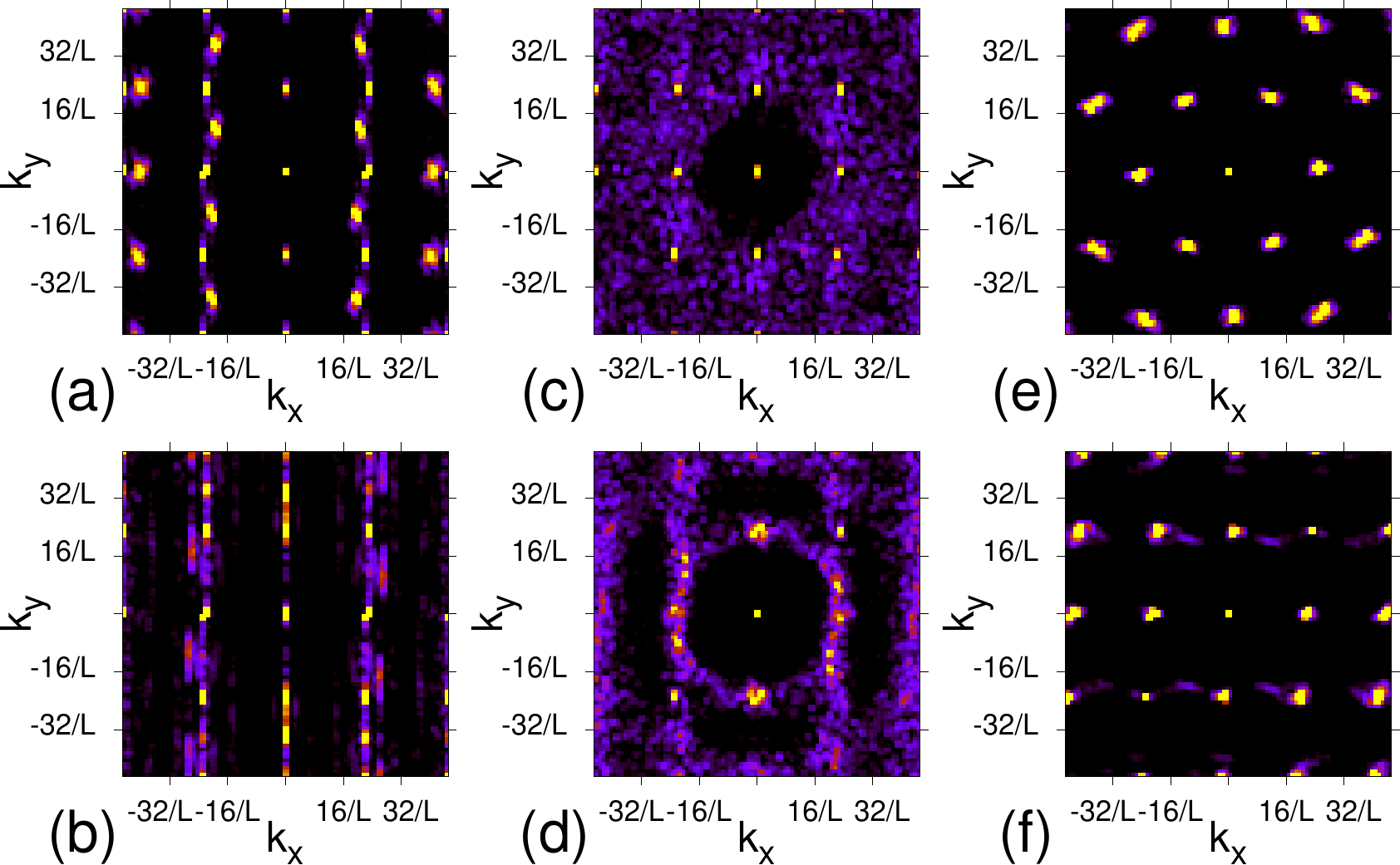}
\caption{ The structure factor $S({\bf q})$ for the system in
  Fig.~\ref{fig:2} with $x$ direction driving at
  $F_p=0.75$ and $\alpha_m/\alpha_d=1.0$.
  (a) Phase I;
  (b) Phase II$_{sb}$;
  (c) Phase III$_{pl}$;
  (d) ML phase;
  (e) MC phase.
  (f) The same system at $F_p=1.0$ has a moving square lattice phase
  described in Section IV.
}
\label{fig:5}
\end{figure}

The dynamic phases are also associated
with changes in the structure factor 
$S({\bf q})$.
In Fig.~\ref{fig:5}(a) we plot $S({\bf q})$ for the system from Fig.~\ref{fig:2}
in phase I at $F_D=0.1$,
where we find a combination of both square and triangular peaks
which
reflects the square ordering of the skyrmions
in the pinned portion of the sample and the triangular ordering
of the skyrmions in the unpinned region.
In phase II$_{sb}$ at $F_D=0.2$ in
Fig.~\ref{fig:5}(b),
the ordering is more smectic
with enhanced peaks
along certain directions due to the sliding of the
skyrmions along the pinning rows.
For $F_D=0.5$ in phase III$_{pl}$, Fig.~\ref{fig:5}(c) shows
that a square ordering has emerged
due to the pinned skyrmions, along with
a smooth background produced by
the liquid
structure of the other skyrmions.
In Fig.~\ref{fig:5}(d) at $F_D=1.0$ in the ML phase,
a ring structure appears along with
six-fold peaks, indicating that there is some short range translational order
imposed on the liquid structure.
In the MC phase at $F_D=1.5$ in
Fig.~\ref{fig:5}(e),
the vortices have triangular ordering and there are six clear peaks.
When the pinning strength is raised to $F_p=1.0$ in the same
system, Fig.~\ref{fig:5}(f) indicates that there is
a moving square lattice,
which we discuss in Section IV.

\begin{figure}
\includegraphics[width=3.5in]{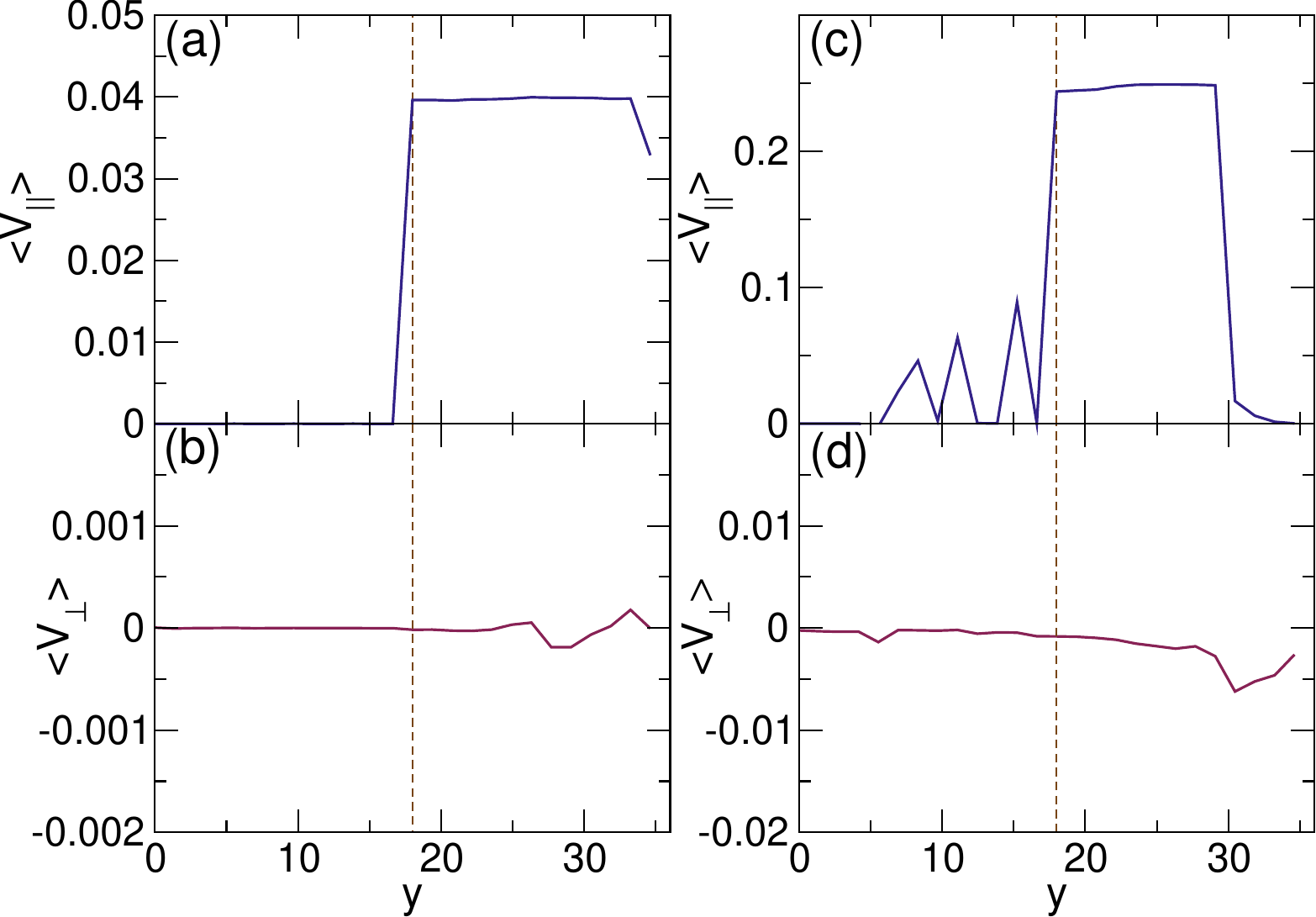}
\caption{The velocities spatially averaged over the $x$ direction
  as a function of $y$ for the
  system in Fig.~\ref{fig:2} with $x$ direction driving at
  $F_p=0.75$ and $\alpha_m/\alpha_d=1.0$.
  (a) $\langle V_{||}\rangle$ and (b) $\langle V_{\perp}\rangle$ in phase I at $F_{D} = 0.04$.
  (c) $\langle V_{||}\rangle$ and (d) $\langle V_{\perp}\rangle$
  in phase II$_{sb}$ at $F_{D} = 0.25$.
  The vertical dashed lines indicate the separation between the pinned
  region ($y\leq 18$) and the pin-free region ($y> 18$).}
\label{fig:6}
\end{figure}

The spatial distribution of the velocities has distinct structures
in the different phases.
In Fig.~\ref{fig:6}(a,b) we plot
$\langle V_{||}\rangle$ and $\langle V_{\perp}\rangle$ averaged over the
$x$ direction as a function of $y$
in phase I at $F_{D} = 0.04$ for the
system in Fig.~\ref{fig:2}.
The vertical dashed line indicates the edge of the pinned region at
$y = 18$.
We find $\langle V_{||}\rangle \approx 0.04$ in the unpinned region 
and $\langle V_{||}\rangle=0$ in the pinned region, 
while $\langle V_{\perp}\rangle$ is zero in both the pinned and unpinned regions. 
As $F_{D}$ increases, the skyrmion density 
in the unpinned region at the largest values of
$y$ decreases since the skyrmion lattice is
being compressed along the $y$-direction.  
In Fig.~\ref{fig:6}(c,d) we show the $x$ direction average values of
$\langle V_{||}\rangle$ and $\langle V_{\perp}\rangle$ versus $y$
in phase II$_{sb}$ at $F_{D} = 0.25$, where
the parallel velocity is finite in the pinned region
and the skyrmion density is close to zero for $y > 35$.
There are  several peaks in $V_{x}$ in the pinned region  that
decrease in height as $y$ decreases, indicating that the motion in the pinned
region is largest close to the edge of the pinning that is exposed to the largest
density of moving skyrmions in the pin-free channel.
Throughout phase II$_{sb}$, $\langle V_{\perp}\rangle = 0$.

\begin{figure}
\includegraphics[width=3.5in]{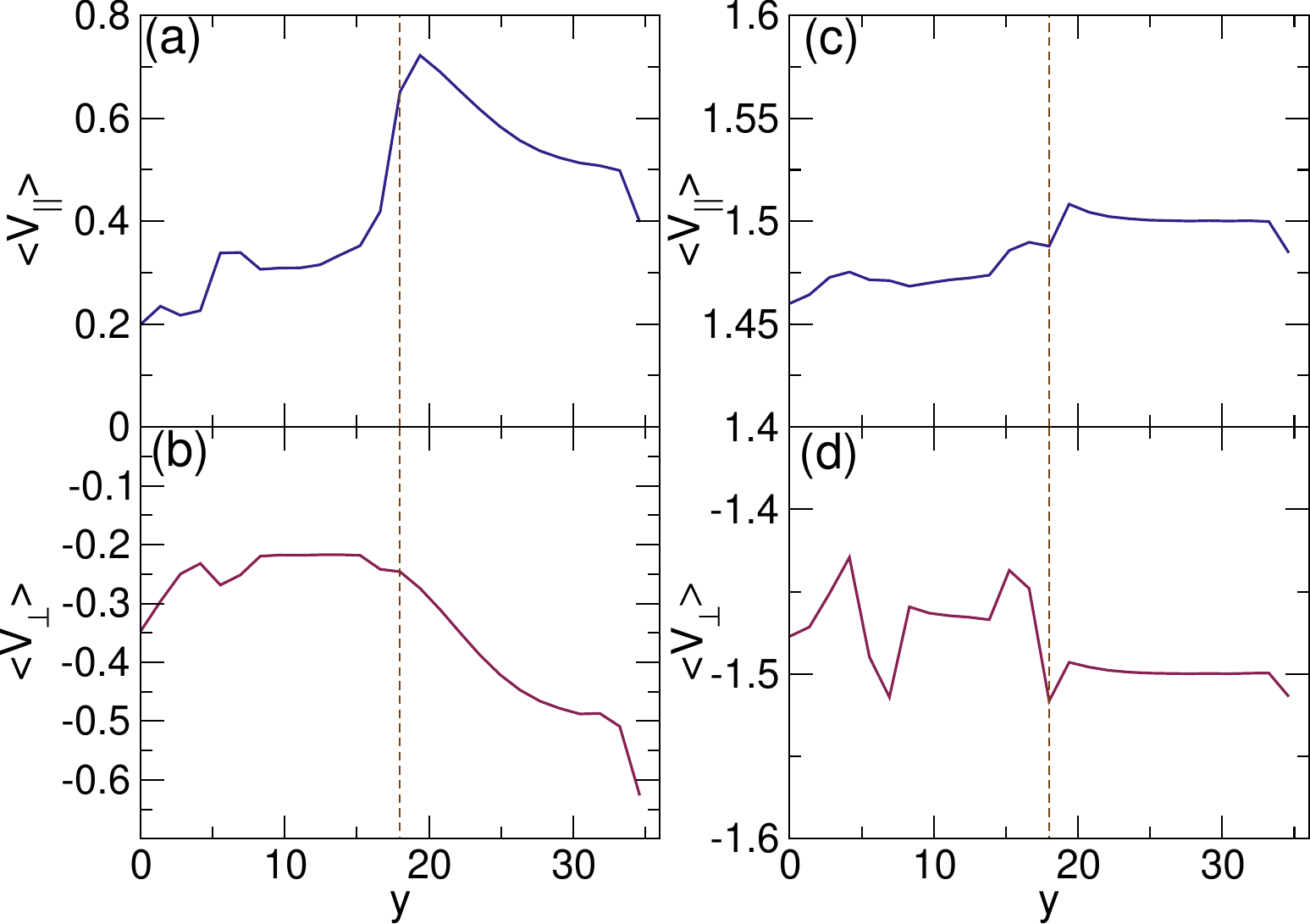}
\caption{ The velocities spatially averaged over the $x$ direction
  as a function of $y$ for the system in Fig.~\ref{fig:2} with $x$ direction
  driving at $F_p=0.75$ and $\alpha_m/\alpha_d=1.0$.
  (a) $\langle V_{||}\rangle$ and (b) $\langle V_{\perp}\rangle$ in phase
  III$_{pl}$ at $F_{D} = 0.5$.
  (c) $\langle V_{||}\rangle$ and (d) $\langle V_{\perp}\rangle$ in the MC phase
  at $F_{D} = 1.5$.  The vertical dashed lines indicate the separation between
  the pinned region ($y\leq 18$) and the pin-free region ($y>18$).
}
\label{fig:7}
\end{figure}

In Fig.~\ref{fig:7}(a,b) we plot the $x$ direction average values of
$\langle V_{||}\rangle$ and $\langle V_{\perp}\rangle$ versus $y$
in phase III$_{pl}$ at $F_{D} = 0.5$.
The velocity is finite both parallel and perpendicular
to the drive, and
the values of both $\langle V_{||}\rangle$
and $\langle V_{\perp}\rangle$ are lowest in the pinned region.
A peak in $\langle V_{||}\rangle$  appears in the unpinned region
near the edge of the pinning
due to a combination of an increase in the skyrmion density with a
speed up or acceleration effect
in which the skyrmions in the pinned region exert a force on the unpinned
skyrmions that is perpendicular to the drive.  This force is rotated into
the driving direction by the Magnus term, contributing an extra velocity
component to the skyrmions in the unpinned region.
A similar acceleration or speed up effect 
was observed
for skyrmions interacting with planar defects
in the form of periodic quasi-one-dimensional potentials \cite{Reichhardt16a}. 
Figure~\ref{fig:6}(c,d) shows the $x$-averaged values of
$\langle V_{||}\rangle$ and $\langle V_{\perp}\rangle$ versus $y$ in
the MC phase at $F_{D} = 1.5$.
Here the velocities in both directions are nearly independent of $y$ since
the effect of the pinning is greatly reduced.  
 
\subsection{Dynamic Phases at Small Magnus Forces}

We next consider the effect of varying the relative strength of the Magnus
force.  We observe three regimes of behavior consisting of the
low Magnus force regime for
$0 \leq \alpha_{m}/\alpha_{d} \leq 0.5$, the intermediate Magnus force regime for
$0.5 < \alpha_{m}/\alpha_{d} < 7.0$, and the high
Magnus force regime for $\alpha_{m}/\alpha_{m} \geq 7.0$.
Each regime has
distinctive dynamics.

\begin{figure}
\includegraphics[width=3.5in]{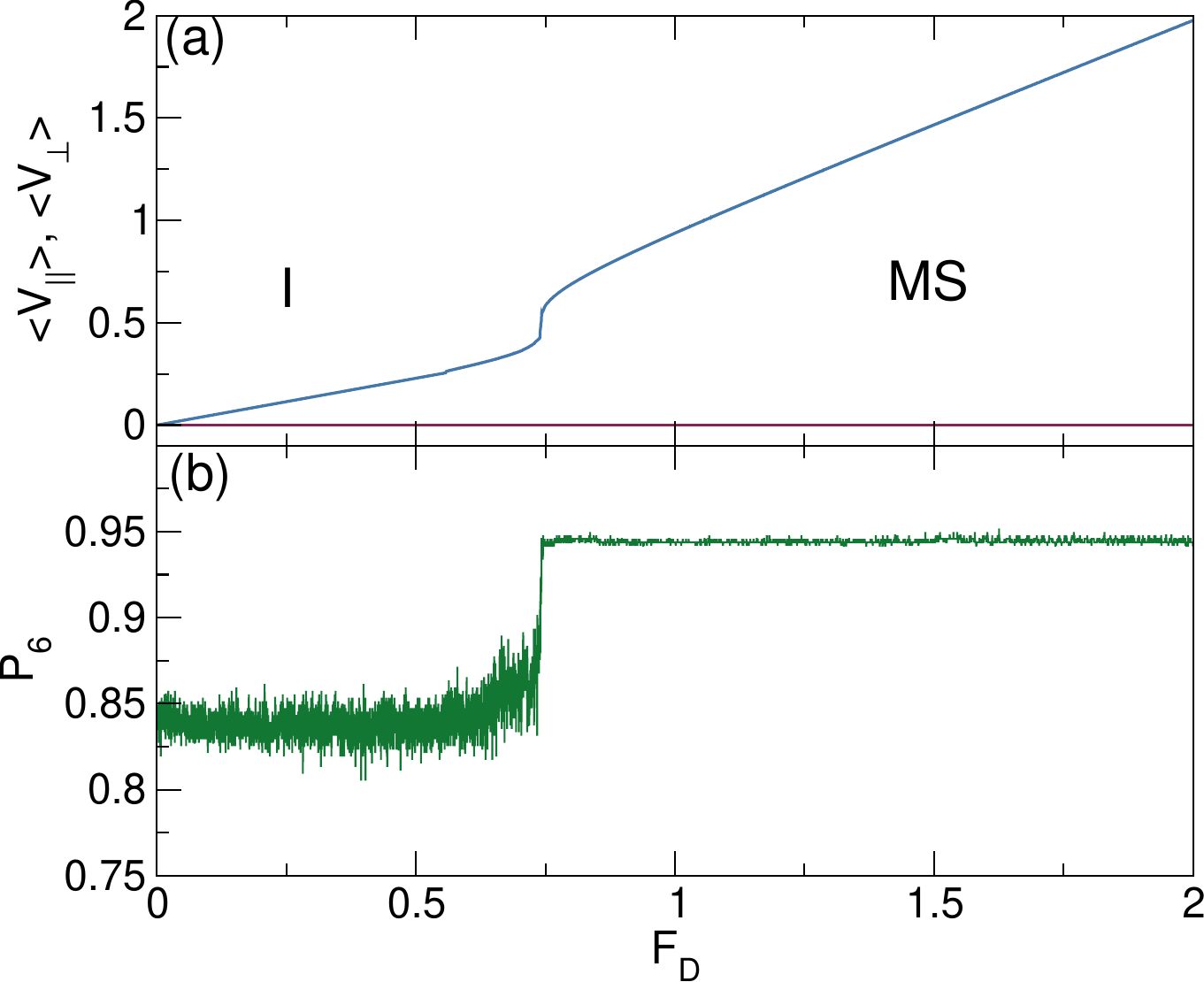}
\caption{(a)
  $\langle V_{||}\rangle$ (blue) and $\langle V_{\perp}\rangle$ (red)
  vs $F_{D}$ for the system in
  Fig.~\ref{fig:2} with $x$ direction driving at 
  $F_{p} = 0.75$ and $\alpha_{m}/\alpha_{d} = 0.$
  We find phase I and
  a moving smectic (MS) phase as illustrated in Fig.~\ref{fig:9}. 
  (b) The corresponding fraction of six-fold coordinated skyrmions
  $P_6$ vs $F_{D}$ showing that in the
MS phase there is a jump up to $P_{6} = 0.95$. } 
\label{fig:8}
\end{figure}

\begin{figure}
\includegraphics[width=3.5in]{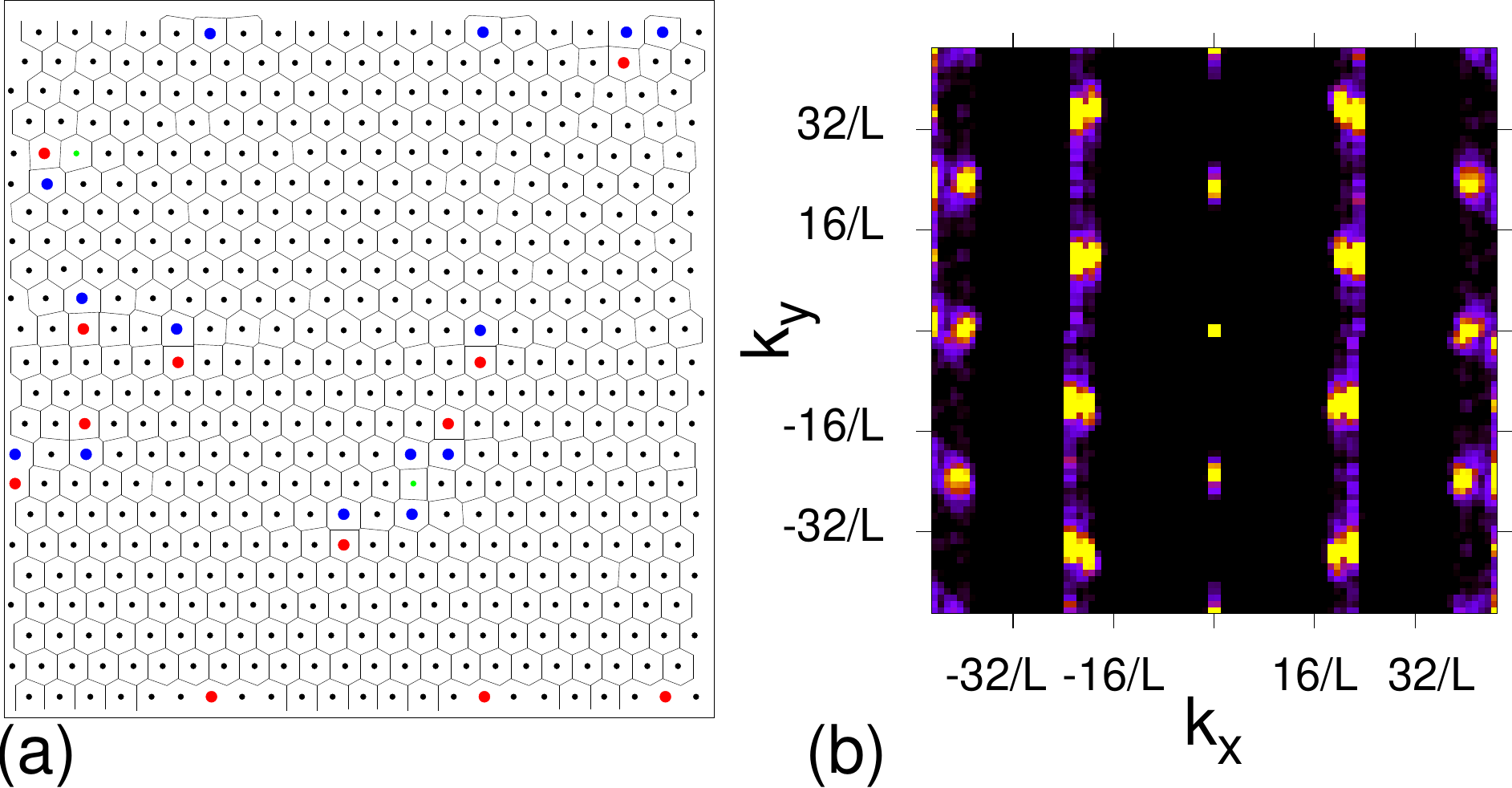}
\caption{ The Voronoi construction for the skyrmion locations
  for the system in Fig.~\ref{fig:8} with $x$ direction driving at 
  $F_p=0.75$ and $\alpha_m/\alpha_d=0$ at $F_{D} = 1.0$ in the
  MS phase.
  The coordination number of individual skyrmions is 5 (red), 6 (black), 7 (blue),
  or outside the range of 5 to 7 (green).
  Fivefold and sevenfold coordinated defects form pairs that have their
  Burgers vector aligned with the driving direction.
  (b) The corresponding structure factor $S({\bf k})$ indicates the presence of
smectic order.}    
\label{fig:9}
\end{figure}

The low Magnus force regime is relevant not only for skyrmion systems but also
for
certain superconducting vortex systems \cite{Ao93}.
In Fig.~\ref{fig:8}(a) we plot
$\langle V_{||}\rangle$ and $\langle V_{\perp}\rangle$ versus $F_{D}$  for 
the system in Fig.~\ref{fig:2}
with $F_{p} = 0.75$ and $\alpha_{m}/\alpha_{m} = 0.$
We find two dynamic phases.
In phase I,
flow occurs only in the unpinned region, while
at higher drives we observe a moving smectic (MS) phase in which
all of the
skyrmions are flowing in both the pinned and unpinned regions. 
In both phases,
$\langle V_{\perp}\rangle = 0.$
We note that 
there can be some shear banding flows near the I-MS transition, which
produce the nonlinear behavior in
$\langle V_{||}\rangle$ near $F_{D}/F_{p} = 1.0$. 
This phenomenon has been discussed
previously in more detail for the overdamped regime \cite{Reichhardt19a}.
Figure~\ref{fig:8}(b) shows the the corresponding $P_{6}$ versus $F_{D}$  
where we observe a jump from  $P_{6} = 0.85$ in phase I to $P_{6} = 0.95$ in the MS phase. 
The value of $P_{6}$ is less than $1.0$
due to pinning-induced dislocations in the skyrmion lattice.
These dislocations cause
the skyrmions in the pinned portion of the 
sample to move a bit more slowly than the skyrmions in the unpinned region,
producing a net slip
between the two spatial regions.
In Fig.~\ref{fig:9}(a) we plot the Voronoi construction for the instantaneous skyrmion
positions
in the MS phase at $F_{D} = 1.0$.
Here, fivefold and sevenfold coordinated defects form pairs with their Burgers vector
aligned in the direction of the drive.
In the corresponding $S({\bf k})$, shown in 
Fig.~\ref{fig:9}(b),
there is strong smectic ordering,
but six peaks are still present due to the partial triangular ordering
of the moving lattice.

\begin{figure}
\includegraphics[width=3.5in]{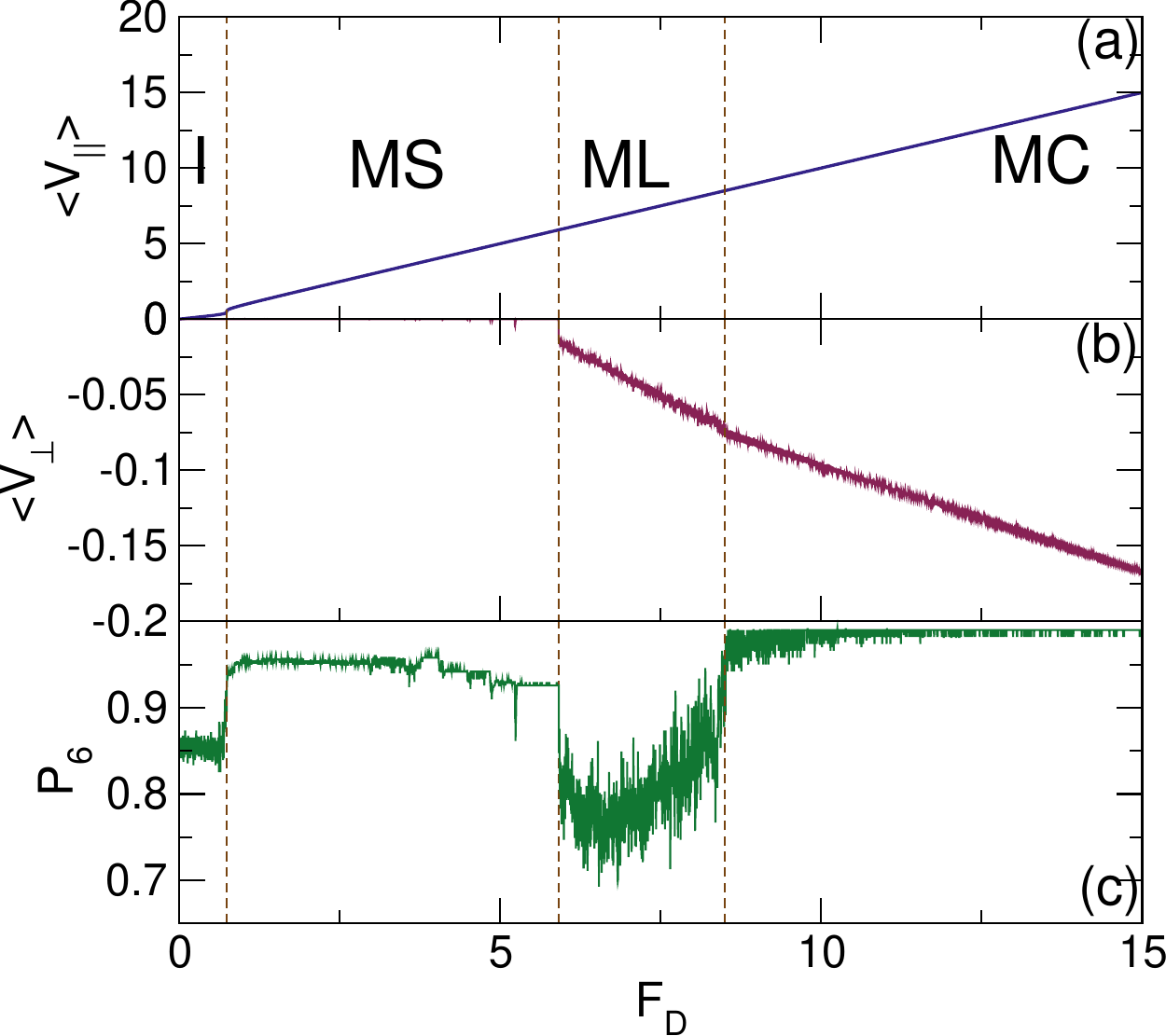}
\caption{ (a) $\langle V_{||}\rangle$,
  (b) $\langle V_{\perp}\rangle$, and
  (c) $P_{6}$ vs $F_{D}$ for the sample 
  in Fig.~\ref{fig:2} with $x$ direction driving at
  $F_p=0.75$ and $\alpha_{m}/\alpha_{d} = 0.0125$.
  The vertical lines distinguish the different dynamic phases.
  I: longitudinal flow in only the pin-free channel;
  MS: moving smectic; ML: moving liquid; MC: moving crystal.
  A drop in $P_6$ marks the window of disordered moving
  liquid between the moving smectic (MS) and moving crystal (MC)
  phases.
} 
\label{fig:10}
\end{figure}

For finite but small $\alpha_{m}/\alpha_{d}$,
we still find
phase I and the MS phase;
however, at higher
drives, an additional transition occurs when
the skyrmion lattice decouples from the pinning, ceases to be locked
in the driving direction, and exhibits
a finite skyrmion Hall angle.
In Fig.~\ref{fig:10}(a,b,c) 
we plot
$\langle V_{||}\rangle$, $\langle V_{\perp}\rangle$, and $P_{6}$,
respectively, versus $F_{D}$ for the same system in Fig.~\ref{fig:8} but at 
$\alpha_{m}/\alpha_{d} = 0.0125$,
where the skyrmions have an
intrinsic Hall angle of $\theta^{\rm int}_{sk} = 0.7125^\circ$. 
The vertical lines distinguish the
different dynamical phases.
We find
extended regions of phase I and the MS phase, in which
$\langle V_{\perp}\rangle = 0$. 
For $5.9 < F_{D} < 8.5$, the system becomes disordered,  
as indicated by the drop in $P_{6}$ to $P_6 \approx 0.75$,
and the system forms a moving liquid.
Here $\langle V_{\perp}\rangle$ becomes finite as the system
begins to show a finite skyrmion Hall angle.  
For $F_{D} > 8.5$, $P_{6}$ jumps up to $P_6=0.985$ when the system enters a
moving crystal phase.
In the MC phase the 
topological ordering is higher than
in the MS since the skyrmion lattice is no longer locked to the square pinning array.  
The MS-MC transition 
is accompanied by a change in the slope of $\langle V_{\perp}\rangle$.

\begin{figure}
\includegraphics[width=3.5in]{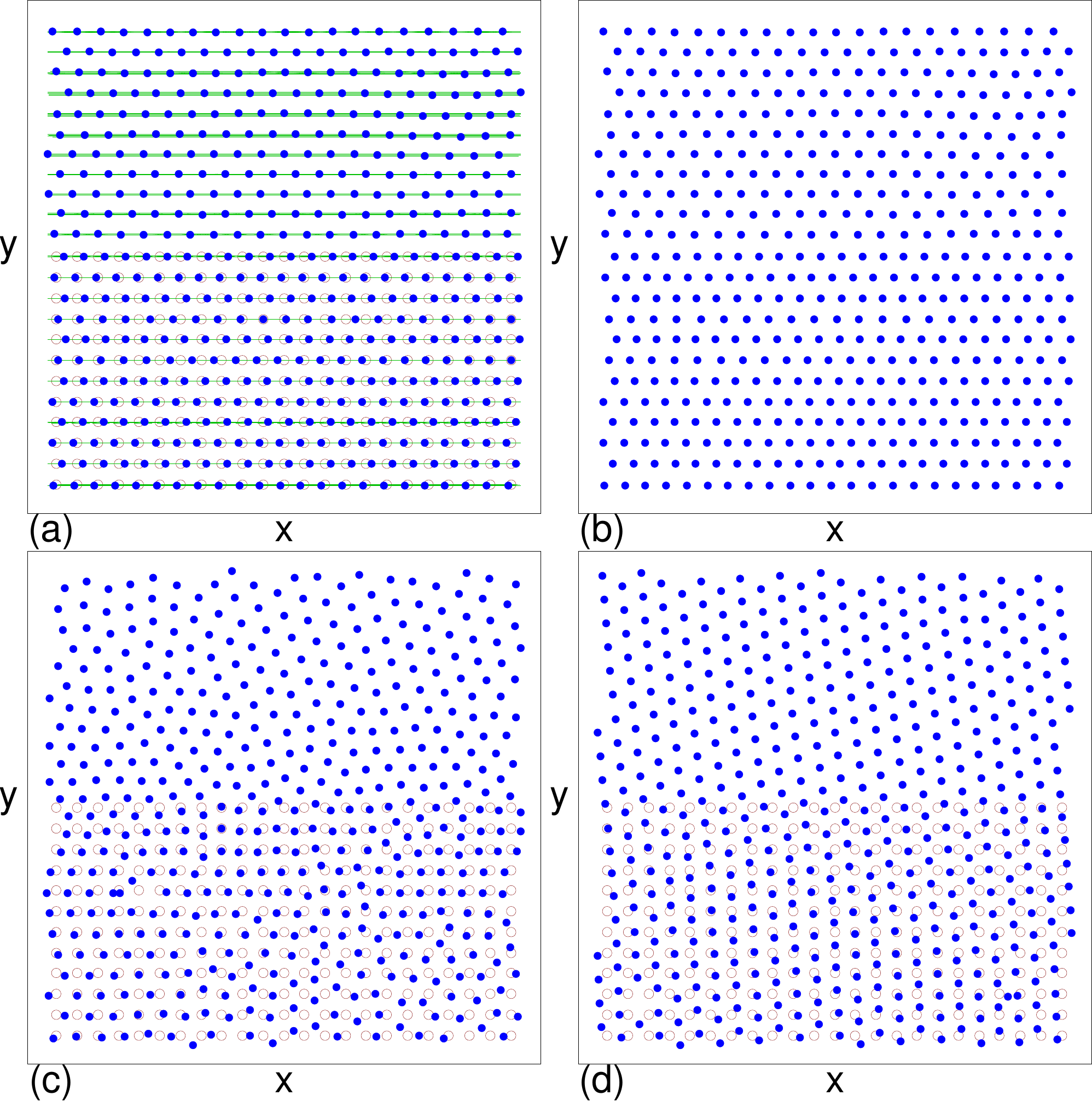}
\caption{
  Skyrmion locations (blue dots) and pinning site locations (open circles)
  for the system in Fig.~\ref{fig:10}
  with $x$ direction driving at $F_p=0.75$ and
  $\alpha_m/\alpha_d=0.0125$.
  (a) The
  MS phase
  at $F_{D} = 2.5$, with skyrmion trajectories
  during a fixed time interval drawn as green lines. 
  (b) The same as panel (a) with only the skyrmion locations shown,
  indicating that
  the skyrmion lattice is aligned in the $x$-direction, parallel to the drive. 
  (c) The
  ML phase at $F_{D} = 7.0$.
  (d) The MC phase at $F_{D} = 7.5$.}  
\label{fig:11}
\end{figure}

In Fig.~\ref{fig:11}(a) we show the skyrmion locations, pinning site locations,
and skyrmion trajectories in the
MS phase for the system in Fig.~\ref{fig:10} at $F_{D}  = 2.5$.
The image in Fig.~\ref{fig:11}(b) of only the skyrmion locations in the same state
indicates that the skyrmion lattice is aligned with the
$x$ axis, parallel to the driving direction.
As $F_D$ increases in the MS phase,
the skyrmions in the pin free region become compressed and row reductions
occur,
producing the drops in $P_{6}$ in the
MS phase at $3.0 < F_D < 5.9$ in Fig.~\ref{fig:10}(c).
At $F_D=7.0$ in the ML phase, illustrated in
Fig.~\ref{fig:11}(c),
the skyrmions in the pinned region tend to remain aligned with
the $x$ axis while the rest of the skyrmions have adopted a rotated configuration.
The formation of a
strongly driven moving liquid state
occurs due to the competition between the pinning, which tends to
lock the skyrmion motion along the $x$ axis,
and the Magnus force,
which favors skyrmion motion at an angle of $7^\circ$ with respect to the
$x$ axis.
At the MS-ML
transition, the dislocations that were gliding
in the MS phase generate
a temporary proliferation of additional topological defects
that dynamically anneal away as $F_D$ increases and
the skyrmions enter the MC phase, illustrated in
Fig.~\ref{fig:11}(d) at $F_D=7.5$.

In general, increasing the drive reduces
the dynamical effect of the pinning,
as found in other systems that exhibit depinning \cite{Reichhardt17},
and eventually the effective pinning strength becomes weak enough that the
skyrmions can start jumping in the direction transverse to the drive.
The guiding
of particles
along a symmetry direction of a periodic pinning array has been well studied in
overdamped systems, where the 
particles tend to lock to symmetry directions of the pinning lattice even
when these directions are not aligned with the direction of the external drive
\cite{Reichhardt17,Reichhardt99,Korda02}. 
This effect has also been studied for 
individual skyrmions moving in periodic pinning arrays
using both a particle-based model \cite{Reichhardt15a} and
continuum-based simulations \cite{Feilhauer19}. 
For skyrmion assemblies,
the locking effect is generally weaker,
but it can be observed for driving over square pinning 
arrays along $0^\circ$ and $45^\circ$ from the major symmetry axis
of the lattice.
The MS-ML and ML-MC transitions shift
to higher values of $F_{D}$ with decreasing
$\alpha_{m}/\alpha_{d}$,
and at $\alpha_{m}/\alpha_{d} = 0$,
the system remains
locked in the MS phase for all drives above depinning.

\begin{figure}
\includegraphics[width=3.5in]{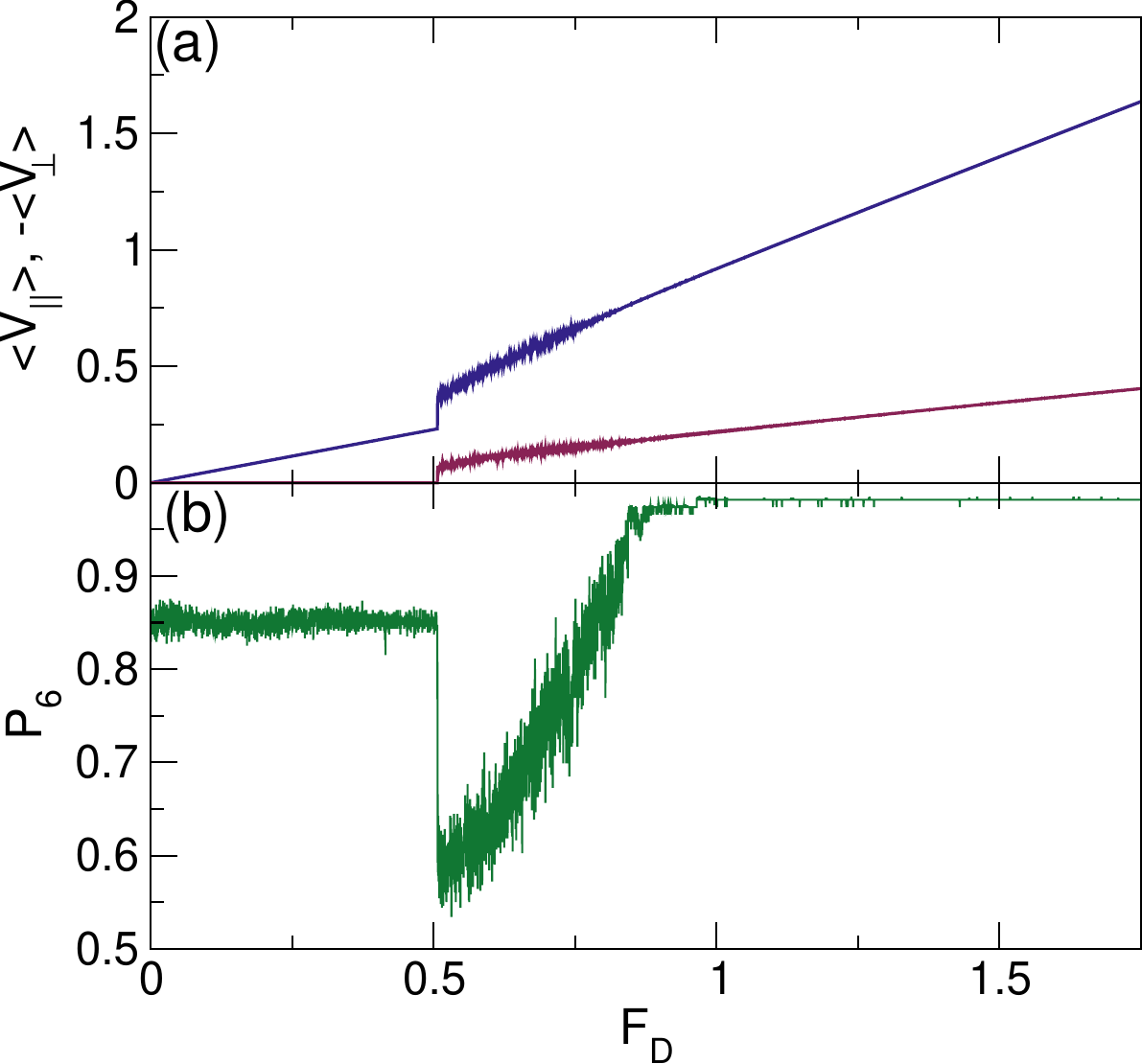}
\caption{(a) $\langle V_{||}\rangle$ (blue) and
  $-\langle V_{\perp}\rangle$ (red) vs $F_{D}$ for the
  system in Fig.~\ref{fig:10} with $x$ direction driving at
  $F_p=0.75$ and $\alpha_{m}/\alpha_{d} = 0.25$. Here
  we have multiplied
  $\langle V_{\perp}\rangle$ by $-1$ for clarity.  
(b) The corresponding $P_{6}$ vs $F_{D}$ showing the absence of the MS phase.  }
\label{fig:12}
\end{figure}

In Fig.~\ref{fig:12}(a) we plot $\langle V_{||}\rangle$ and
$-\langle V_{\perp}\rangle$ versus $F_{D}$ for the same system in Fig.~\ref{fig:10} at
a higher value of $\alpha_{m}/\alpha_{d} = 0.25$,
which corresponds to $\theta^{\rm int}_{sk} = 14^\circ$.
The corresponding $P_6$ versus $F_D$ in
Fig.~\ref{fig:12}(b)
shows a transition from phase I
to the plastic flow phase III$_{pl}$ followed by a transition
directly into the moving crystal phase,
with no MS phase.

\begin{figure}
\includegraphics[width=3.5in]{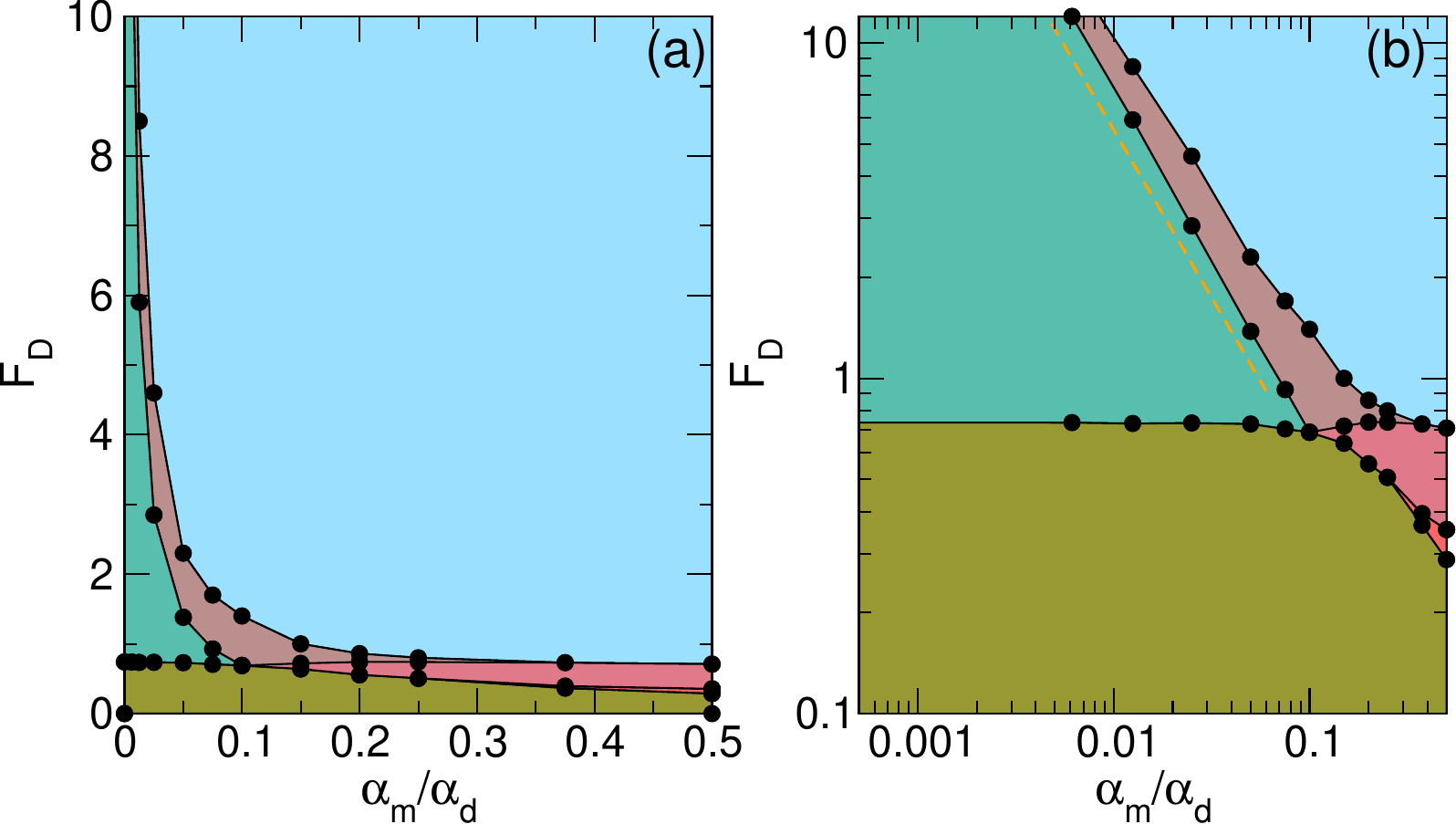}
\caption{ The dynamic phase diagram as a function of $F_{D}$ vs
  $\alpha_{m}/\alpha_{d}$ for the low Magnus force regime.
  Phase I: olive green;
  moving smectic (MS) phase: teal;
  moving liquid (ML) phase: brown;
  plastic flow phase III$_{pl}$: pink;
  shear band phase II$_{sb}$: red;
  and the moving crystal (MC) phase: light blue.
  The MS phase occurs only when $\alpha_{m}/\alpha_{d} < 0.1$.
  (b) The same data on a log-log scale.
  The dashed orange line is a fit to $F_D \propto 1/(\alpha_m/\alpha_d)$,
  showing that the drives at which the MS-ML and ML-MC phase transitions occur
diverge as $\alpha_m/\alpha_d$ decreases.
}
\label{fig:13}
\end{figure}

We construct a dynamic phase regime as a function of
$F_{D}$ versus $\alpha_{m}/\alpha_{d}$ in Fig.~\ref{fig:13}(a)
for the low Magnus force systems in 
Figs.~\ref{fig:10} to \ref{fig:12}.
The MS phase only occurs when $\alpha_{m}/\alpha_{d} < 0.1$.
In Fig.~\ref{fig:13}(b) we replot the phase diagram
on a log-log scale.
The dashed line is a fit to $F_{D}\propto 1/(\alpha_m/\alpha_d)$ indicating that
the drives at which
the MS-ML and ML-MC transitions occur
diverges as the relative strength of the Magnus force decreases.
An interesting aspect of this result
is that it
suggests that in a system such as superconducting vortices with a small but
finite Magnus force,
there could be a transition from a moving smectic to a moving crystal state at
a finite but large drive.

\subsection{Intermediate and High Magnus Force}

\begin{figure}
\includegraphics[width=3.5in]{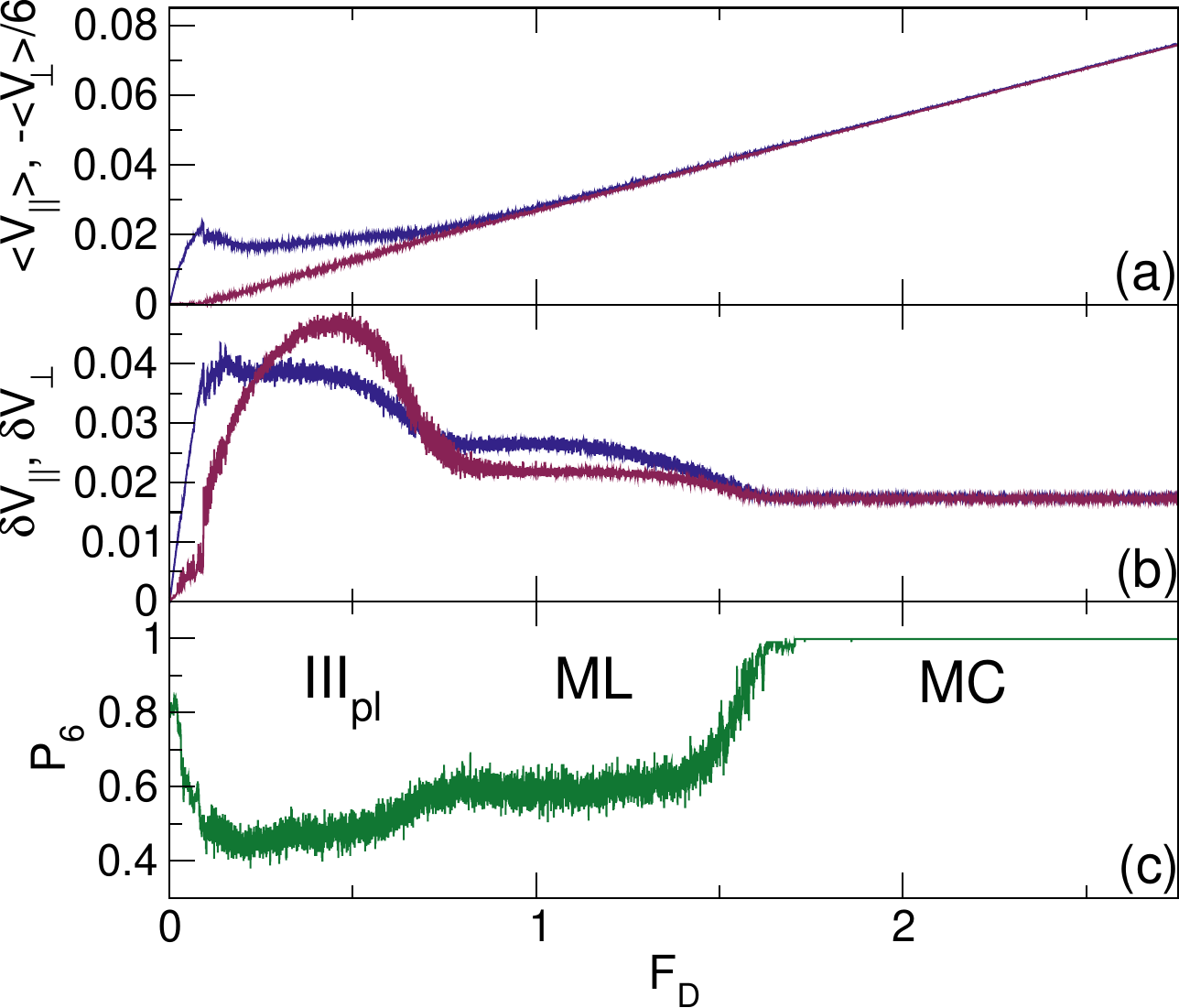}
\caption{  
  (a) $\langle V_{||}\rangle$ (blue) and $-\langle V_{\perp}\rangle/6$ (red) vs $F_{D}$ for
  a system with $x$ direction driving at $F_p=0.75$ and
  $\alpha_{m}/\alpha_{d} = 6.0$.
  For clarity, we have multiplied $\langle V_{\perp}\rangle$ by $-1$
  and divided it by the value of $\alpha_m/\alpha_d$
  in order to normalize $\langle V_{\perp}\rangle$ to
  the value of $\langle V_{||}\rangle$ at high drives.
  (b) The corresponding velocity deviations $\delta V_{||}$ (blue) and
  $\delta V_{\perp}$ (red) vs $F_D$.
  (c) The corresponding $P_{6}$ vs $F_{D}$, where we find an extended window
  of moving liquid (ML) phase. III$_{pl}$: plastic flow phase; MC: moving crystal phase.}
\label{fig:14}
\end{figure}

In the intermediate Magnus force regime of
$0.5 < \alpha_{m}/\alpha_{d} < 7.0$,
we find the same five phases I, II$_{sb}$, III$_{pl}$, ML, and MC described above,
with an expansion of the ML phase since
the
driving force at which
the ML-MC transition occurs increases
with increasing Magnus force.
We plot $\langle V_{||}\rangle$ and $\langle V_{\perp}\rangle$ versus $F_D$
in Fig.~\ref{fig:14}(a)
for a system with $\alpha_{m}/\alpha_{d} = 6.0$. 
For clarity, we have normalized
$\langle V_{\perp}\rangle$ to the value
of $\langle V_{||}\rangle$ at high drives by
dividing $\langle V_{\perp}\rangle$ by $\alpha_m/\alpha_d$ and
multiplying it by $-1$.
We can also characterize the dynamic phases
by measuring the skyrmion velocity deviations in the $x$ and $y$-directions 
as in previous work \cite{Reichhardt18,Reichhardt19g}, 
$\delta V_{||} = \sqrt{[\sum^{N}_{i}(v^{i}_{||})^2 - \langle V_{||}\rangle^2]/N}$ and
$\delta V_{\perp} = \sqrt{[\sum^{N}_{i}(v^{i}_{\perp})^2 - \langle V_{\perp}\rangle^2]/N}$.
The plot of $\delta V_{||}$ and $\delta V_{\perp}$ versus $F_D$ in
Fig.~\ref{fig:14}(b) indicates that the velocity deviations are largest in
the plastic flow phase, and diminish to a constant value in the moving crystal phase.
In Fig.~\ref{fig:14}(c) we show $P_6$ 
versus $F_{D}$ for the same system.
When $F_D/F_p>1.0$, 
we find
an extensive region of ML phase
in which $P_{6}$ has a higher value than in the plastic flow phase but
a lower value than in the moving crystal phase.

\begin{figure}
\includegraphics[width=3.5in]{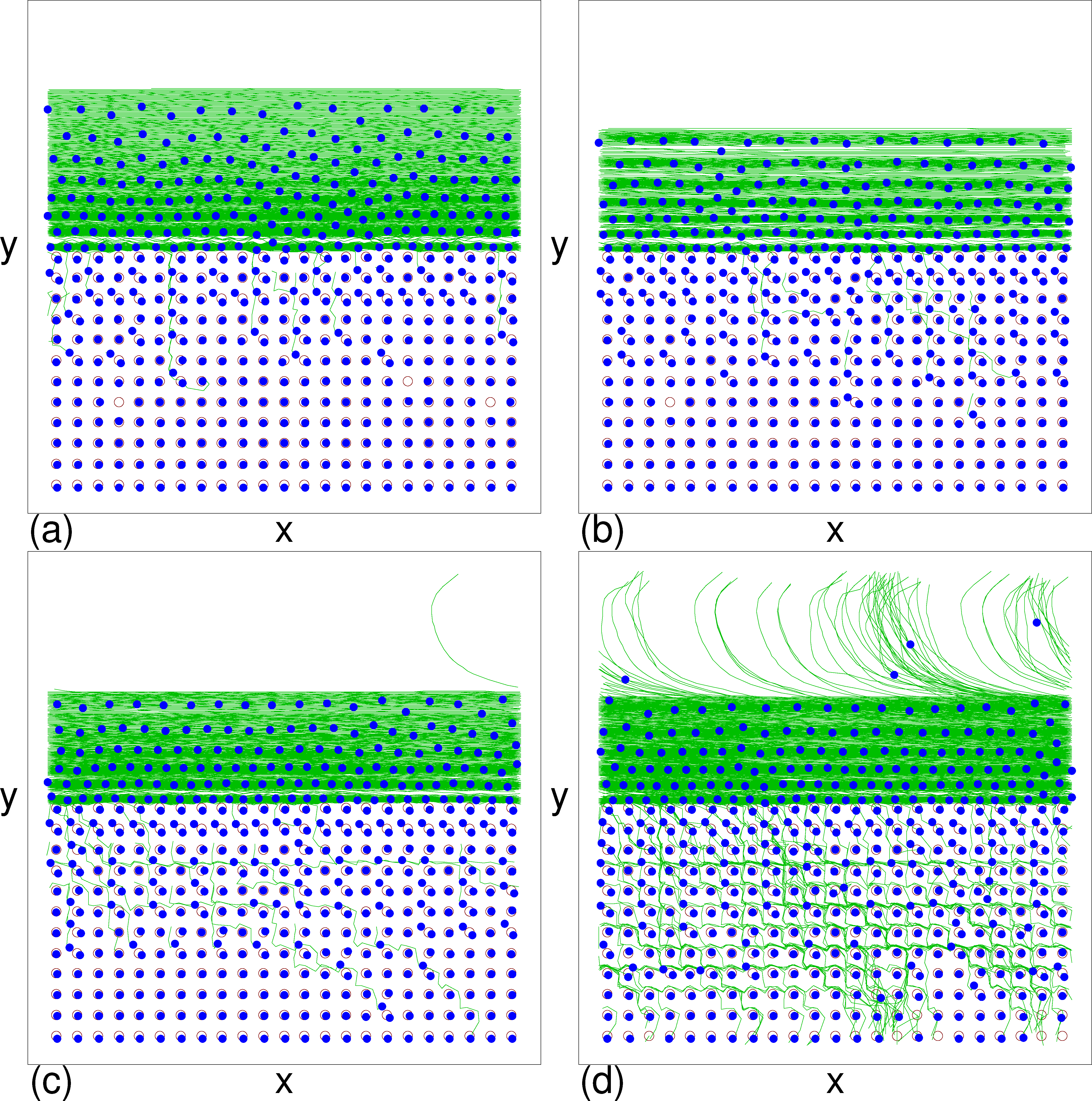}
\caption{
  Skyrmion locations (blue dots), pinning site locations (open circles), and skyrmion
  trajectories (green lines) during a fixed time interval for
  a system
  with $x$ direction driving at $F_p=0.75$ and
  $\alpha_m/\alpha=2.5$, where
  the shear banding phase is replaced with
  the avalanche phase $II_{a}$.
  Moving skyrmions from the unpinned region
  enter the pinned region
  in the form of avalanches at
  (a) $F_{D} = 0.1$, (b) $F_{D} = 0.2$, and (c) $F_{D} = 0.3$.
  (d) The plastic flow phase III$_{pl}$ at $F_D=0.5$.  }    
\label{fig:15}
\end{figure}

The shear banding phase II$_{sb}$ disappears when
$\alpha_{m}/\alpha_{d} > 2.0$, and it is
replaced by phase II$_a$ in which skyrmions enter the pinned region
via avalanches but where there is almost no flow parallel to the drive.
In Fig.~\ref{fig:15}
we illustrate a system with  $\alpha_{m}/\alpha_{d} = 2.5$ at
$F_{D} = 0.1$, 0.2, 0.3, and $0.5$.
The motion in the pinned region
at $F_{D} = 0.1$ and $0.2$
in Fig.~\ref{fig:15}(a,b) is at almost $90^\circ$ to the drive, 
while for $F_D=0.3$ in Fig.~\ref{fig:15}(c) there is some motion at a
lower angle. 
This effect is similar to what occurs in the Bean state
found in superconductors, where vortices enter from the edge of the sample,
creating a gradient in the vortex density \cite{Bean64,Reichhardt95}.
In the skyrmion case, even though the skyrmion density builds up at the edge of
the pinned region, the strong Magnus force suppresses motion in the $x$ direction,
parallel to the drive.
Eventually when $F_{D}$ is large enough,
skyrmions can travel all the way across the pinned region
and the system enters the plastic flow phase III$_{pl}$
as shown in Fig.~\ref{fig:15}(d) at $F_D=0.5$.
In this work we do not characterize the statistics of the avalanches in
phase II$_a$;
however, previous
work on skyrmion avalanches for density gradient driven skyrmions
analyzed the avalanches in terms of 
a critical phenomenon \cite{Reichhardt18},
and since we have a similar density gradient
in our system,
we expect that the avalanche statistics should be similar.

\begin{figure}
\includegraphics[width=3.5in]{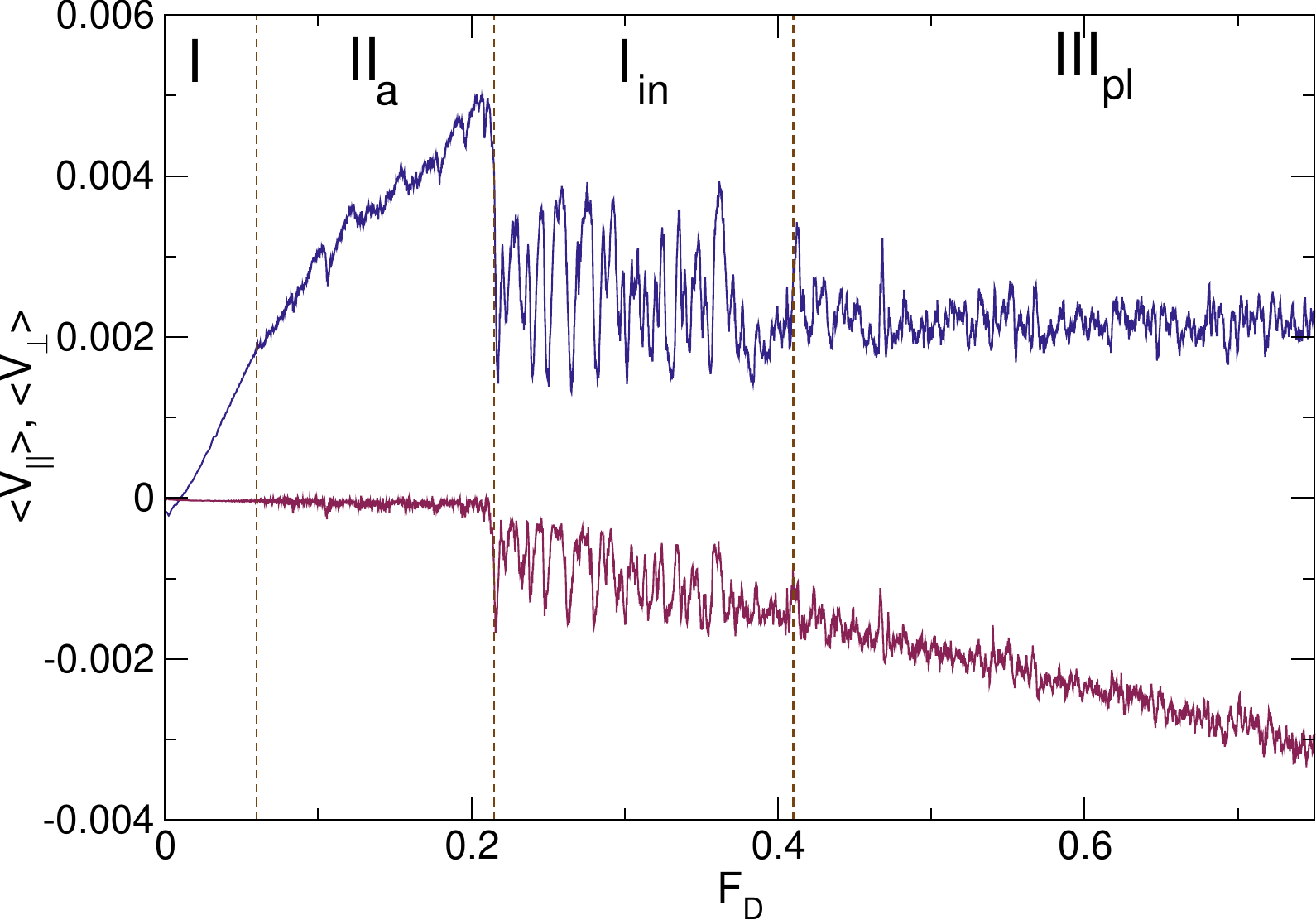}
\caption{ $\langle V_{||}\rangle$ (blue) and $\langle V_{\perp}\rangle$ (red)
  vs $F_{D}$ in a sample with $x$ direction driving at
  $F_p=0.75$ and $\alpha_{m}/\alpha_{d} = 10$.
  Above phase II$_{a}$, there
  is a region of large oscillations
  in the intermittent phase I$_{in}$,
  where
  large scale
phase separated flow occurs as shown in Fig.~\ref{fig:17}.}     
\label{fig:16}
\end{figure}

\begin{figure}
\includegraphics[width=3.5in]{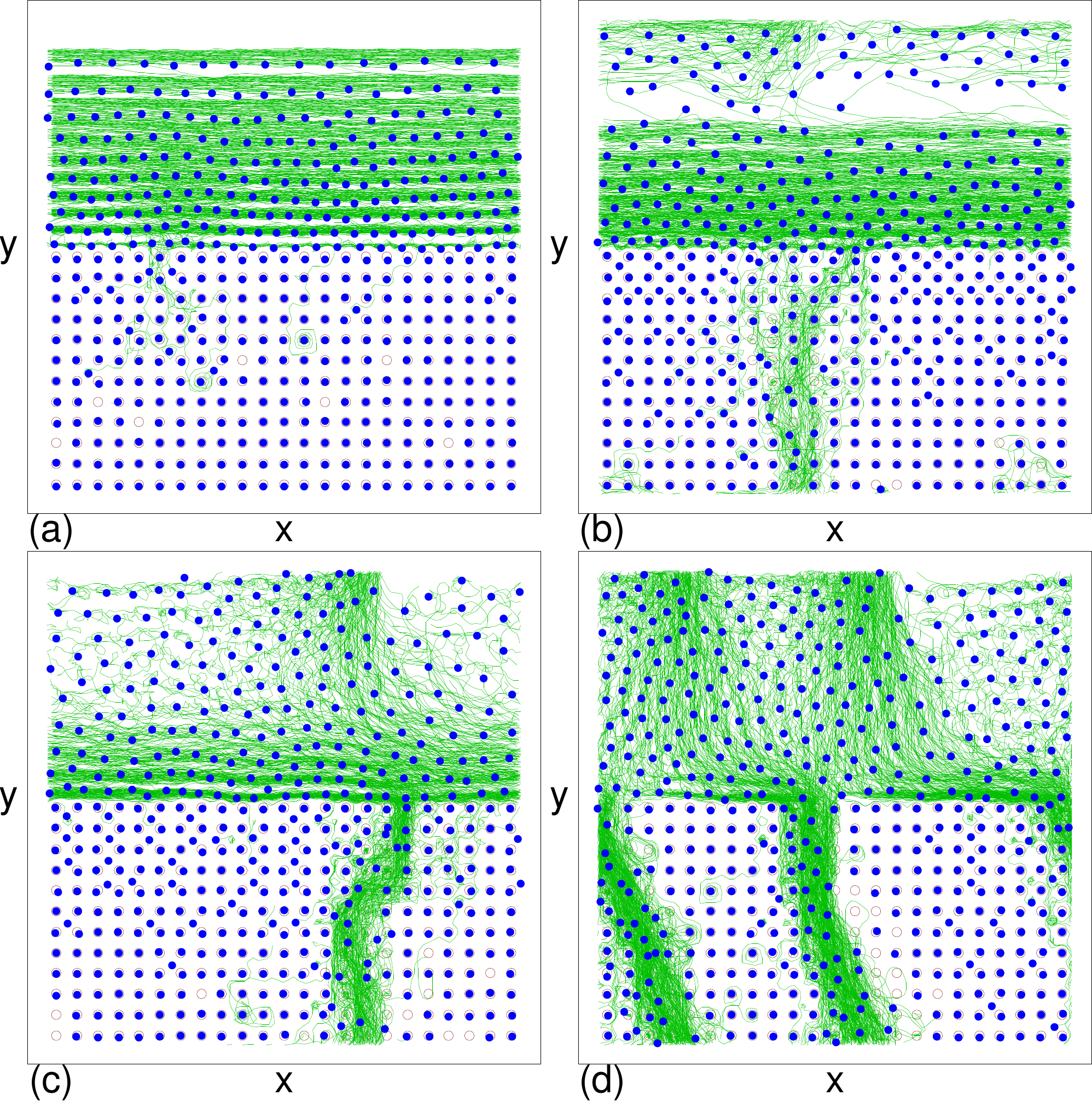}
\caption{
  Skyrmion locations (blue dots), pinning site locations (open circles), and skyrmion
  trajectories (green lines) during a fixed time interval for the system in Fig.~\ref{fig:16}
  with $x$ direction driving at $F_p=0.75$ and $\alpha_m/\alpha_d=10$
  in phase I$_{in}$.
(a) $F_{D} = 0.12$, (b) $F_{D} = 0.25$, (c) $F_{D} = 0.3185$ and (d) $F_{D} = 0.325$.} 
\label{fig:17}
\end{figure}

For $\alpha_{m}/\alpha_{d} > 7.0$ we observe a new phenomenon
in which a portion of the plastic flow phase develops
strong intermittency,
with skyrmion flow occurring in large scale phase separated
moving 
channels in the pinned region.
These channels  
change in shape and size as a function of time,
producing strong oscillations in the velocity.
In Fig.~\ref{fig:16} we plot
$\langle V_{||}\rangle$
and $\langle V_{\perp}\rangle$
versus $F_D$ for
a sample with $\alpha_{m}/\alpha_{d} = 10$.
Here,
after the initial phase II$_{a}$,
in phase I$_{in}$
there are
strong oscillations in the velocity both parallel and perpendicular to the drive
which 
decrease in amplitude until the sample enters phase III$_{pl}$ at
$F_{D} = 0.42$.
In Fig.~\ref{fig:17} we highlight the
skyrmion motion in phase I$_{in}$.
At $F_D=0.12$,
Fig.~\ref{fig:17}(a) shows that a channel has formed
along which the skyrmions enter the pinned region at 
nearly $90^\circ$ to the driving direction.
Several instances of spiraling
motion of individual skyrmions appear,
which are indicative of the
large value of the Magnus force.
In Fig.~\ref{fig:17}(b)
the same system at $F_{D} = 0.25$ 
contains a phase
separated flow
confined to a single river moving at
$-90^\circ$ to the driving direction through
the pinned region.
At $F_D=0.3185$ and 0.325 in Figs.~\ref{fig:17}(c,d),
the large scale rivers
change position as function of both time and drive,
and the large spike in $\langle V_{||}\rangle$
in Fig.~\ref{fig:16} corresponds to two  or three
flowing rivers, while the minimum corresponds to one flowing river.
As $F_{D}$ increases, more of these rivers begin to flow,
and when the system enters phase III$_{pl}$,
the flow is disordered but uniform. 
Phase segregation dynamics
has been observed previously in the particle based model when 
both $\alpha_{m}/\alpha_{d}$ and the pinning
strength are large \cite{Reichhardt18}.
Continuum based simulations also show that
dynamical  
segregation effects can occur when the pinning
strength is sufficiently large \cite{Koshibae18}. 
When the drive is large enough, the effect of the pinning
is reduced and the dynamical behavior becomes more uniform.
We observe the intermittent phase I$_{in}$
for $\alpha_{m}/\alpha_{d}$ as high as $30$, which
is the largest value we considered.
In general, the phase separated dynamics 
or intermittency occurs only when $F_{D}/F_{p} < 1.0$.
The segregation occurs since when the Magnus force is large,
skyrmions that are close together tend to spiral around one another
rather than moving apart. In regions where the
skyrmion density is largest,
the pinning effectiveness is reduced,
leading to additional flow in the same location.     

\begin{figure}
\includegraphics[width=3.5in]{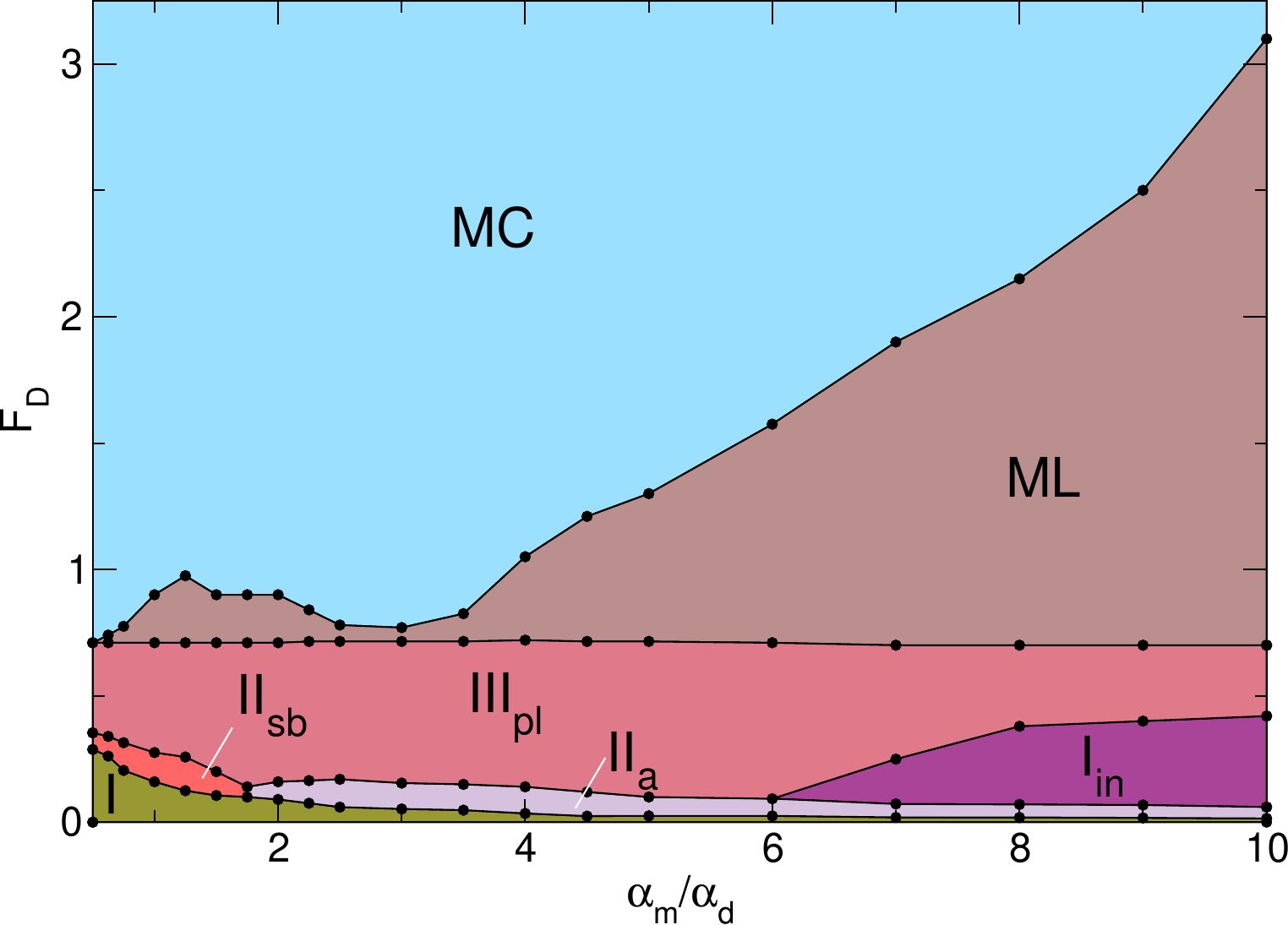}
\caption{
  The dynamic phase diagram as a function of $F_{D}$ vs $\alpha_{m}/\alpha_{d}$
  in samples with $x$ direction driving at $F_p=0.75$,
  showing phases I (olive green),
  II$_{sb}$ (red),
  III$_{pl}$ (pink),
  II$_{a}$ (light purple),
  I$_{in}$ (violet),
  ML (brown), 
  and MC (blue).
  Here phase II$_{a}$ appears when
  $\alpha_{m}/\alpha_{d} > 1.9$ and
  phase I$_{in}$ appears when $\alpha_{m}/\alpha_{d} \geq 7.0$.}  
\label{fig:18}
\end{figure}

In Fig.~\ref{fig:18} we highlight
the dynamic phase diagram as a function of $F_D$ versus $\alpha_m/\alpha_d$
for the intermediate and high Magnus force regimes of
$0.5 \leq \alpha_{m}/\alpha_{d} \leq 10$.
Here the shear banding phase II$_{sb}$ occurs for $0.5 < \alpha_{m}/\alpha_{d} \leq 1.9$,
but is replaced by phase II$_{a}$ for $\alpha_{m}/\alpha_{d} > 1.9$.
When  $\alpha_{m}/\alpha_{d} > 7.0$,
a window of phase I$_{in}$ appears between phases II$_{a}$ and III$_{pl}$. 
The driving force at which the
ML-MC transition occurs increases with increasing $\alpha_{m}/\alpha_{d}$
when $\alpha_m/\alpha_d>3$.
The increase in the
width of the ML phase as $\alpha_m/\alpha_d$ increases
occurs since higher values of
$\alpha_{m}$
produce more
motion associated with the non-dissipative Magnus force,
which creates spiraling motion of the skyrmions when they interact
with the pinning sites.

\begin{figure}
\includegraphics[width=3.5in]{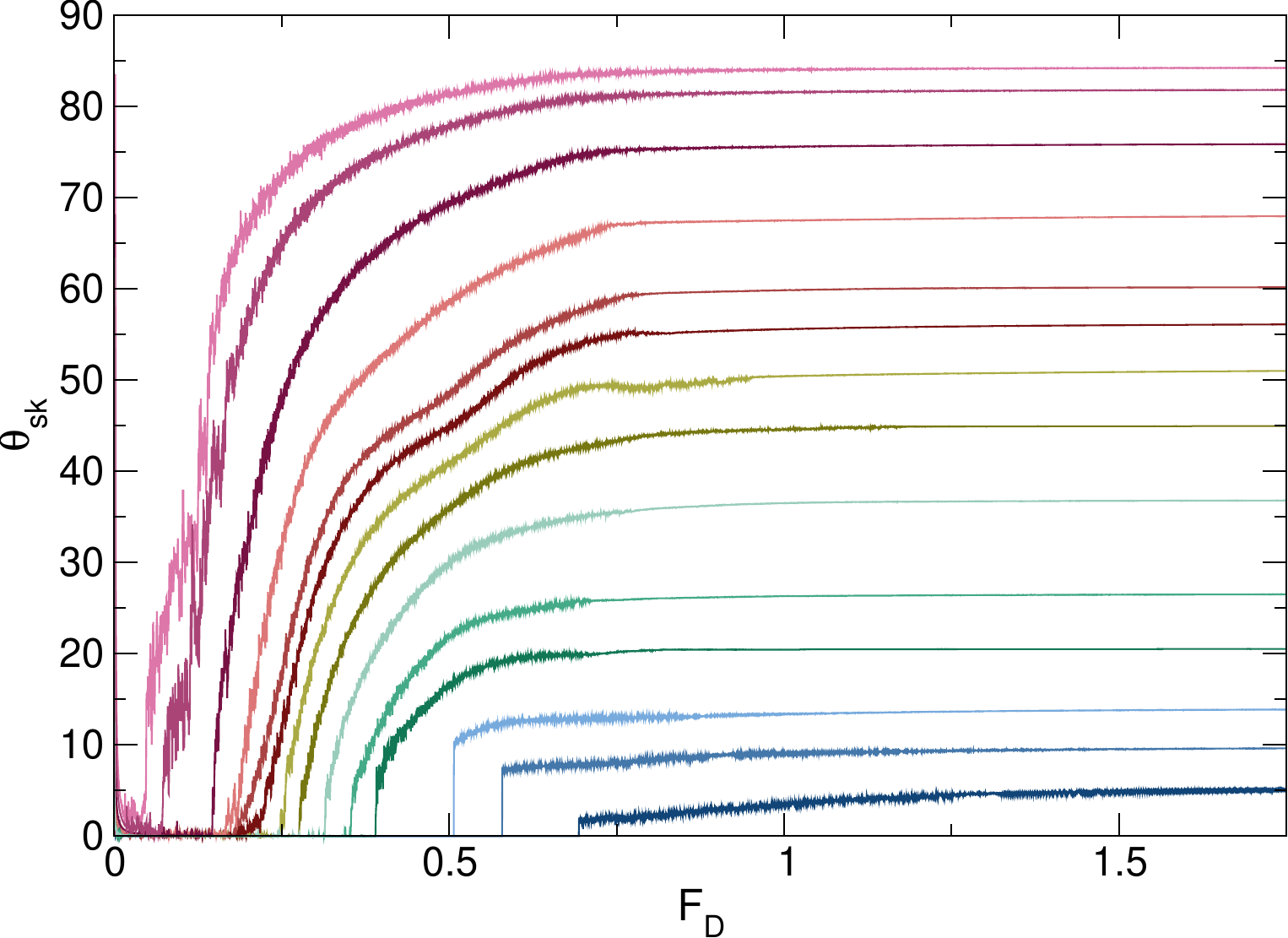}
\caption{ The evolution of $\theta_{sk}$ vs $F_{D}$ for
  samples with $x$ direction driving at $F_p=0.75$ and
  $\alpha_{m}/\alpha_{d} = 10$, 7, 5,
  2.5, 1.75, 1.25, 1.0,
  0.75, 0.7, 0.5,
  0.375, 0.25, 0.175,
  and $0.125$,
from top to bottom.} 
\label{fig:19}
\end{figure}

In Fig.~\ref{fig:19} we show the evolution of the skyrmion Hall angle
$\theta_{sk}$ versus $F_D$ for varied $\alpha_{m}/\alpha_{d}$ ranging from 0.125 to 10.
In each case, $\theta_{sk}$ starts at zero when $F_D=0$
and increases with increasing $F_D$
in the plastic flow phase.
There is a smaller increase in $\theta_{sk}$ with increasing $F_D$
in the ML phase, and $\theta_{sk}$ reaches
a saturation value in the MC phase.
Another interesting effect is that
$\theta_{sk}$ changes discontinuously when
$\theta_{sk} < 15^\circ$.

\section{Varied Pinning Strength}

We next consider the effect of varying the pinning strength $F_p$ while holding
the ratio of the Magnus force to the damping term fixed at
$\alpha_{m}/\alpha_{d} = 1.0$.
The behavior can be divided into
three regimes depending on whether the pinning
is weak, intermediate, or strong.

\begin{figure}
\includegraphics[width=3.5in]{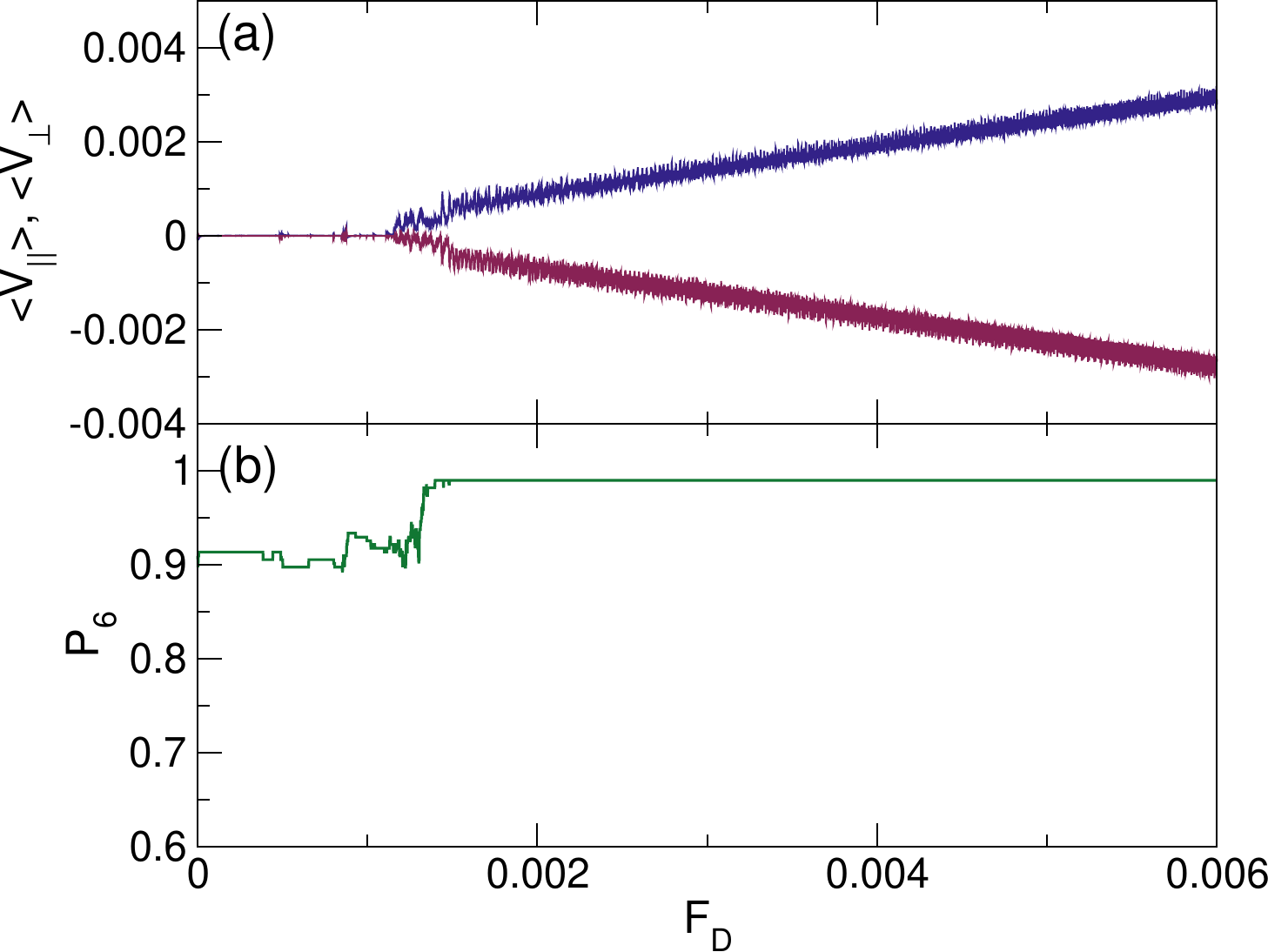}
\caption{ A system with
  $x$ direction driving at 
  $\alpha_{m}/\alpha_{d} = 1.0$ and $F_{p} = 0.01$,
  where an elastic depinning transition in which the skyrmions keep their same
  neighbors separates the pinned phase P from the MC phase.
  (a) $\langle V_{||}\rangle$ (blue) and $\langle V_{\perp}\rangle$ (red) vs $F_{D}$. 
(b) The corresponding $P_{6}$ vs $F_{D}$.}
\label{fig:20}
\end{figure}

\begin{figure}
\includegraphics[width=3.5in]{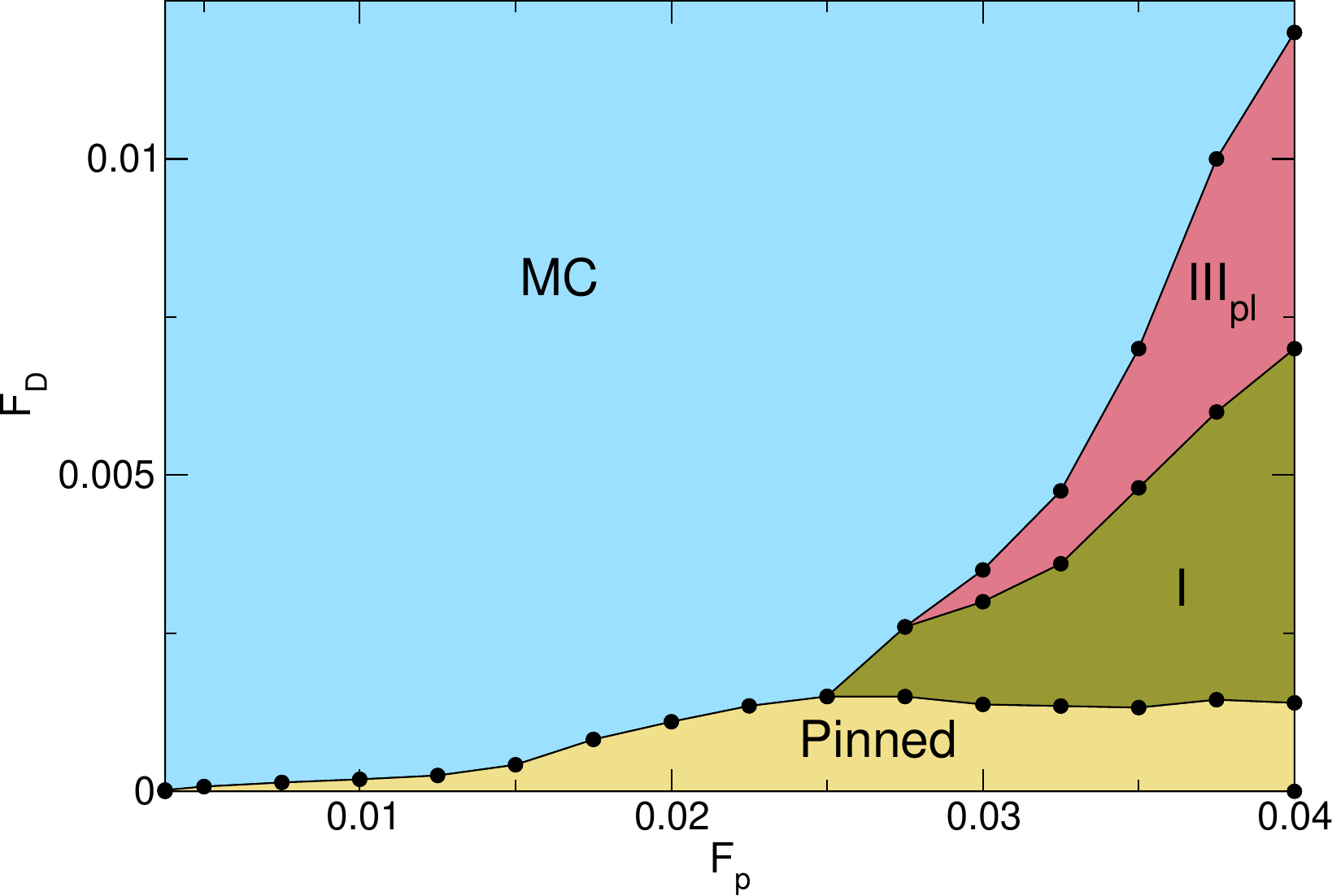}
\caption{Dynamic phase diagram
  as a function of $F_D$ vs $F_p$ in the weak pinning regime
  for samples with $x$ direction driving and
  $\alpha_m/\alpha_d=1.0$, showing the
  pinned phase (yellow),
  moving crystal MC (light blue),
  plastic flow III$_{pl}$ (pink),
  and phase I (olive green).}
\label{fig:21}
\end{figure}

The weak pinning regime
appears when $F_{p} < 0.05$.
Here the motion is elastic
and each skyrmion keeps its same neighbors for all values of $F_{D}$.
At low drives we find
a pinned phase $P$ in which the skyrmions in the pinned region are able to prevent
the skyrmions in the unpinned region from moving due to the
skyrmion-skyrmion interaction forces.
As the drive increases,
an elastic depinning transition occurs
into a moving lattice phase with a finite skyrmion Hall angle,
as shown in Fig.~\ref{fig:20}(a)
where we plot $\langle V_{||}\rangle$ and $\langle V_{\perp}\rangle$ versus $F_D$. 
Figure~\ref{fig:20}(b) shows the corresponding $P_{6}$
versus $F_D$ curve, which changes only slightly at the depinning transition
from the coexisting pinned squares and hexagonal lattice
into a moving lattice.
When the pinning is even weaker,
there is a transition in the pinned state from a pinned square lattice to
a floating triangular lattice,
and the depinning threshold drops even further.  
For $F_{p} > 0.03$, 
we observe a separate depinning transition of
the skyrmions in the unpinned region
which move at zero Hall angle,
follows by a plastic flow region
and a moving crystal regime.
In Fig.~\ref{fig:21} we show the dynamic phase diagram in the weak pinning regime
as a function of $F_D$ versus $F_p$ where we
highlight the pinned phase P, moving crystal MC,
phase I motion, and the plastic flow state III$_{pl}$.

\begin{figure}
\includegraphics[width=3.5in]{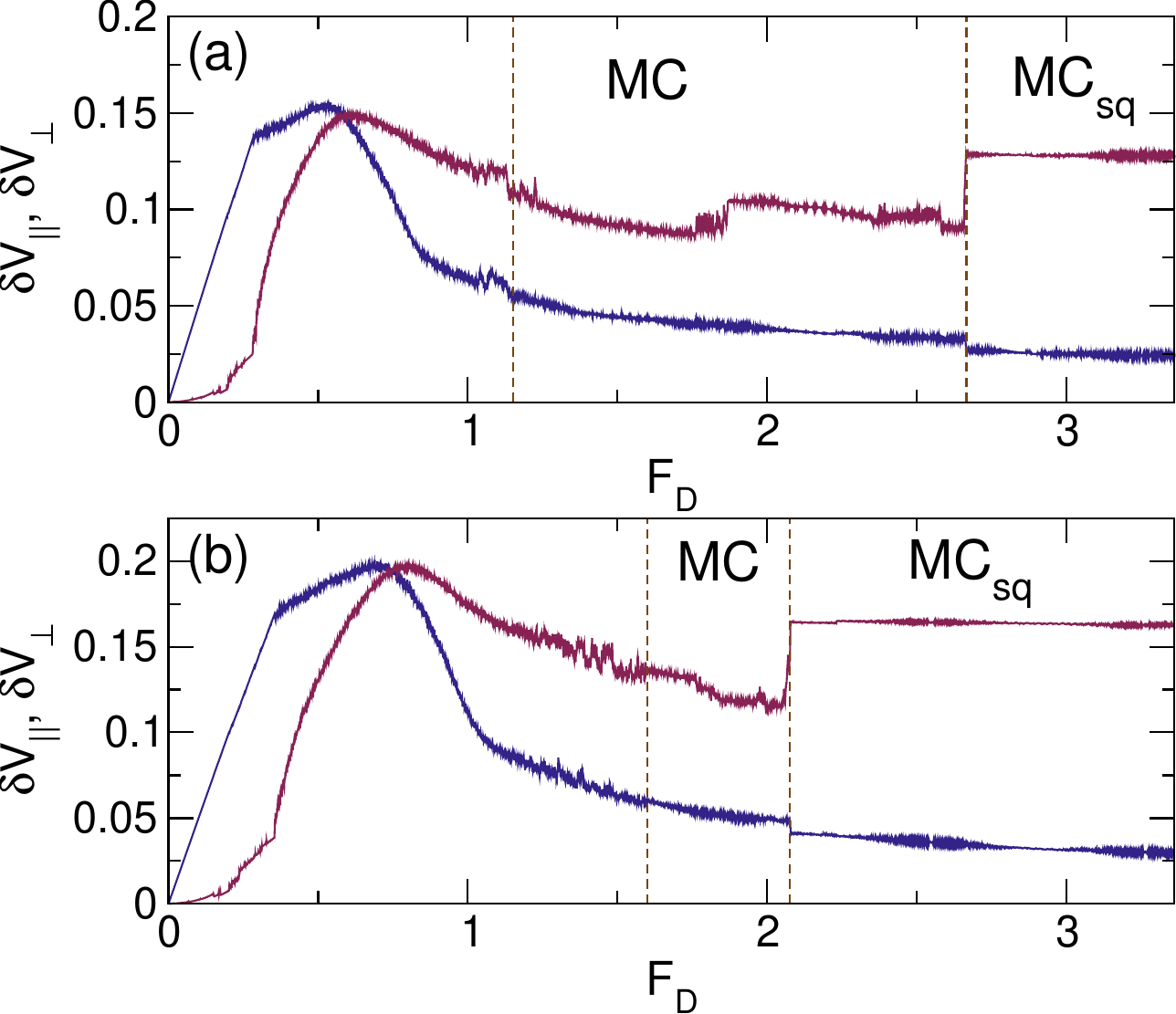}
\caption{Velocity deviations $\delta V_{||}$ (blue) and $\delta V_{\perp}$ (red) vs $F_{D}$
  for samples with $x$ direction driving and
  $\alpha_{m}/\alpha_{d} = 1.0$ showing a transition
  from a triangular moving crystal MC to a
  square moving crystal MC$_{sq}$.
  (a) $F_p = 0.8$. (b) $F_{p} = 1.0$. }
\label{fig:22}
\end{figure}

In the intermediate pinning regime $0.05 < F_{p} < 3.5$,
we observe phases I and III$_{pl}$ as well as the MC state.
At $\alpha_{m}/\alpha_{d} = 1.0$, the intrinsic skyrmion Hall angle is
$\theta_{sk}^{\rm int}=45^\circ$,
which corresponds to a locking direction of the square pinning array in the pinned
portion of the sample.
For $F_{p} < 0.85$, the elastic energy associated with
the triangular skyrmion lattice in the MC state is large enough
that the skyrmion motion
does not lock with the pinning lattice;
however, as the pinning strength increases, we find a
regime in which
the pinning can induce a  directional locking, as indicated by
the formation of a square moving skyrmion lattice
of the type shown
in Fig.~\ref{fig:5}(f).  
In Fig.~\ref{fig:22}(a) we plot $\delta V_{||}$ and $\delta V_{\perp}$
versus $F_D$ for a sample with $F_{p} = 0.8$, 
highlighting the formation of the triangular MC phase
followed by a transition to the square moving crystal phase MC$_{sq}$.
Within the MC phase, occasional small rotations of the triangular lattice
occur which produce
the jumps in in $\delta V_{\perp}$.
At $F_p=1.0$, the $\delta V_{||}$ and $\delta V_{\perp}$ versus $F_D$ curves in
Fig.~\ref{fig:22}(b) indicate that
the transition to the MC$_{sq}$
state has shifted to
lower drives.
For smaller values of $F_{p}$,
the MC phase disappears and there is a transition
directly from
a ML phase to the MC$_{sq}$ phase.

\begin{figure}
\includegraphics[width=3.5in]{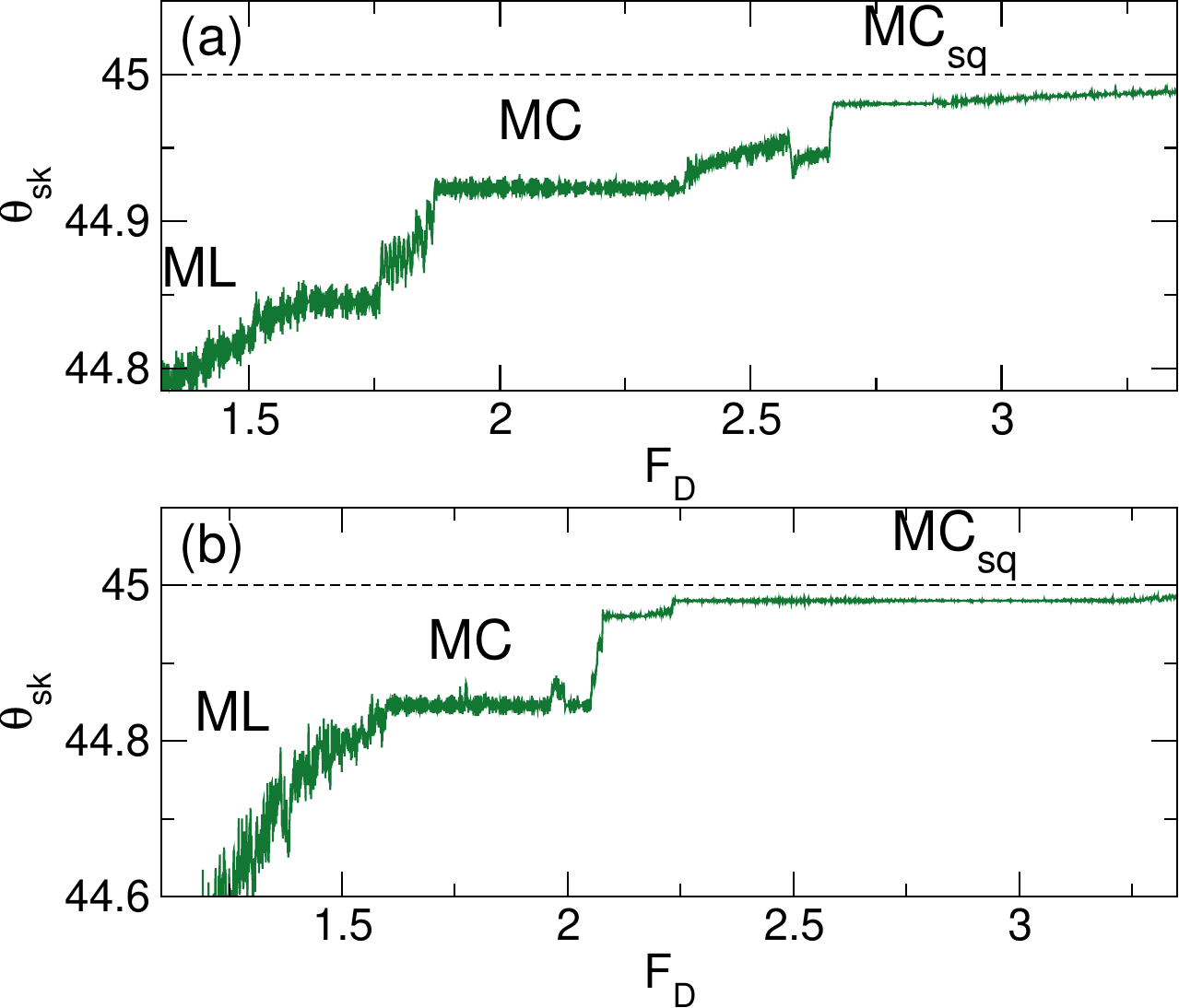}
\caption{$\theta_{sk}$ vs $F_D$ showing
the evolution of the skyrmion Hall angle across the ML-MC-MC$_{sq}$ transitions
for the system in Fig.~\ref{fig:21}
with $x$ direction driving and
$\alpha_m/\alpha_d=1.0$.
(a) $F_{p} = 0.8$. (b) $F_{p} = 1.0$. }
\label{fig:23}
\end{figure}

In Fig.~\ref{fig:23}(a,b) we show a blowup of
the skyrmion Hall angle $\theta_{sk}$ versus $F_D$ across
the ML-MC-MC$_{sq}$ transitions for the system in Fig.~\ref{fig:22}
at $F_p=0.8$ and $F_p=1.0$, respectively.
Here, $\theta_{sk}$ is lower in the
MC phase
than in the MC$_{sq}$ phase.
At $F_p=0.8$ in Fig.~\ref{fig:23}(a),
the onset of the MC phase coincides with a flattening
of $\theta_{sk}$,
but as $F_D$ increases, several jumps in $\theta_{sk}$ occur
due to large scale rotations
of the moving lattice,
and finally in the MC$_{sq}$ state the jumps in $\theta_{sk}$ disappear.
The angle of motion
in the MC$_{sq}$ is slightly smaller than $45^\circ$
due to a weak
guiding effect of the skyrmions along the boundaries separating the
pinned and unpinned regions of the sample.

\begin{figure}
\includegraphics[width=3.5in]{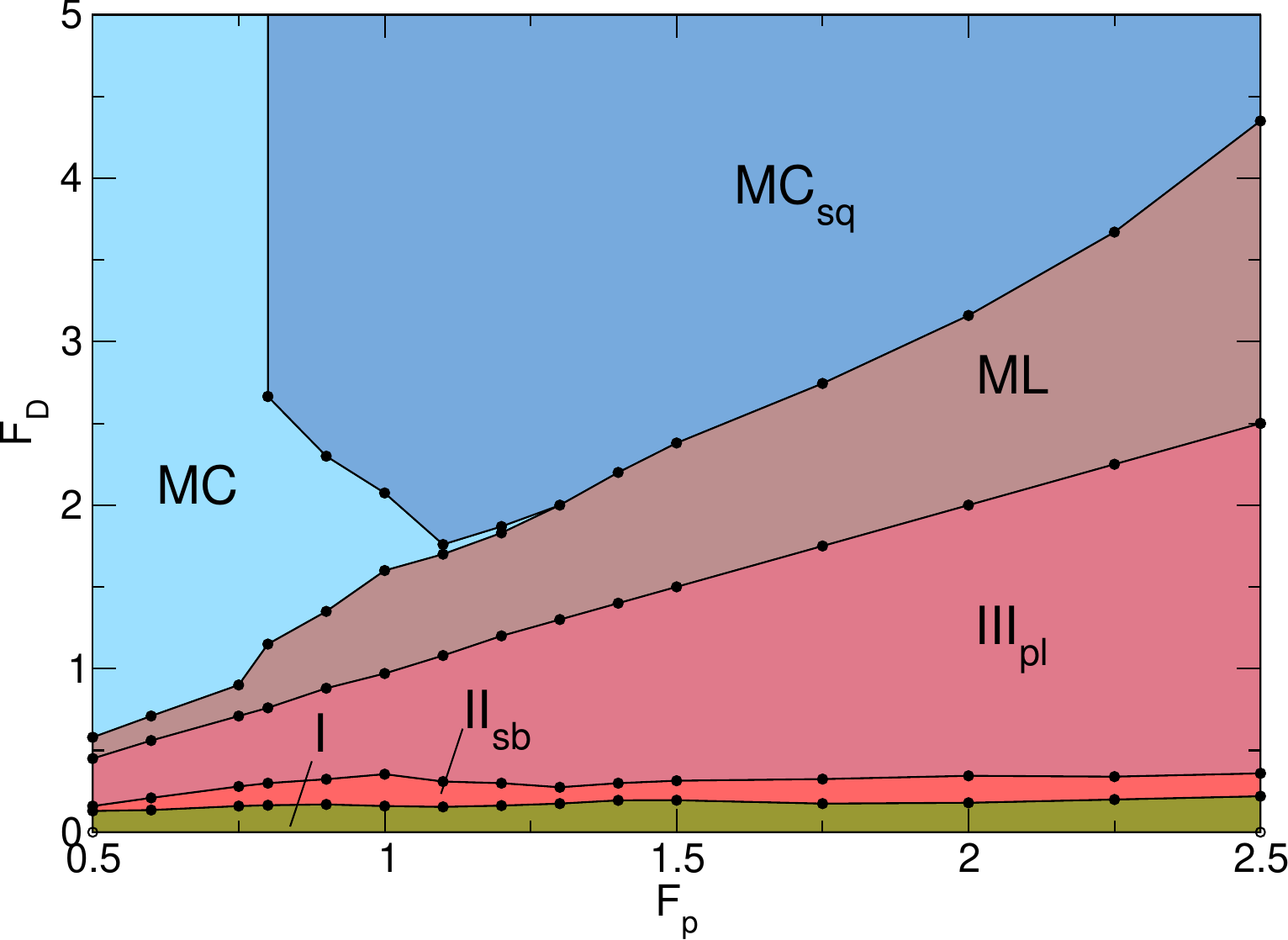}
\caption{
  Dynamic phase diagram
  as a function of $F_D$ vs $F_p$ for the intermediate pinning regime
  in samples with $x$ direction driving and $\alpha_m/\alpha_d=1.0$
  showing phases I (olive green), II$_{sb}$ (red), III$_{pl}$ (pink),
  ML (brown), MC (light blue), and MC$_{sq}$ (dark blue).  
}
\label{fig:24}
\end{figure}

In Fig.~\ref{fig:24} we construct a dynamic phase diagram as a function of $F_D$
versus $F_p$
for the intermediate pinning regime,
highlighting phases I, II$_{sb}$, III$_{pl}$, ML, MC, and MC$_{sq}$.
We note that there
is still a pinned phase P at small $F_{D}$; however, on the scale
of the figure this phase cannot be seen. 
The MC-MC$_{sq}$ transition
shifts to higher values of $F_{D}$ as the
pinning strength is lowered.
We have also examined varied $\alpha_{m}/\alpha_{d}$ for 
different pinning strengths and find similar phases
to those shown in Fig.~\ref{fig:24};
however, the MC$_{sq}$ phase appears only when $\alpha_{m}/\alpha_{d}$
is near 1.0.
An intermittent phase I$_{in}$ appears
for low values of $\alpha_m/\alpha_d$ when $F_p$ is large.

\begin{figure}
\includegraphics[width=3.5in]{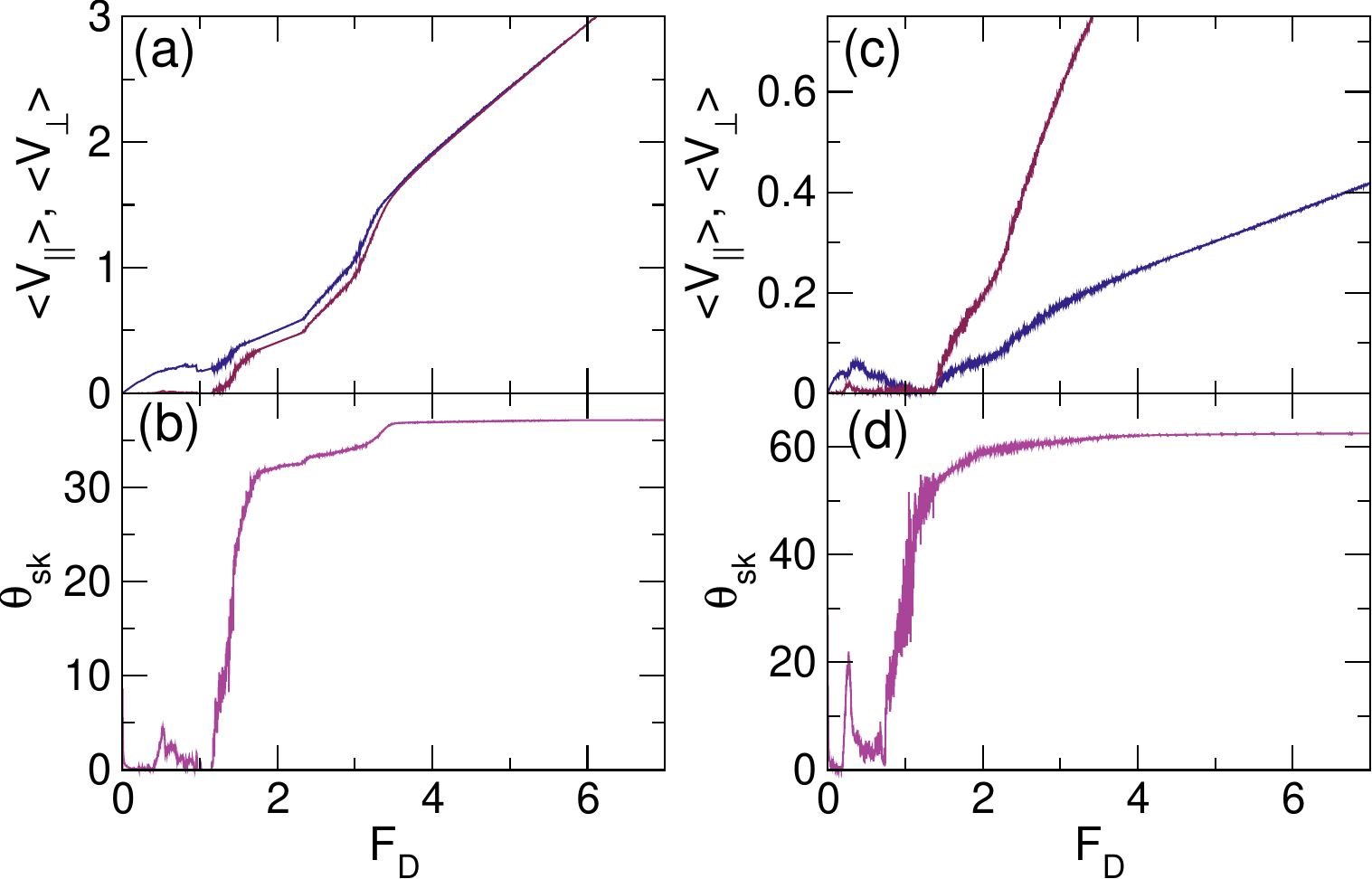}
  \caption{ (a) $\langle V_{||}\rangle$ (blue) and $\langle V_{\perp}\rangle$ (red)
    vs $F_{D}$ for a sample with $x$ direction driving, 
 $\alpha_{m}/\alpha_{d} = 1.0$, and $F_{p} = 4.5$ showing regions in which
    $d\langle V_{||}\rangle/dF_{D} < 0.0$,
    also known as negative differential mobility,
    due to the trapping of multiple skyrmions by individual pinning sites.
    (b) The corresponding $\theta_{sk}$ vs $F_D$
    where multiple regimes appear.
    (c)  $\langle V_{||}\rangle$ and $\langle V_{\perp}\rangle$ vs $F_{D}$
    for a sample with $x$ direction driving, 
    $\alpha_{m}/\alpha_{d} = 4.0$, and $F_{p} = 4.5$. 
(d) The corresponding $\theta_{sk}$ vs $F_{D}$.  
}
\label{fig:25}
\end{figure}

\begin{figure}
\includegraphics[width=3.5in]{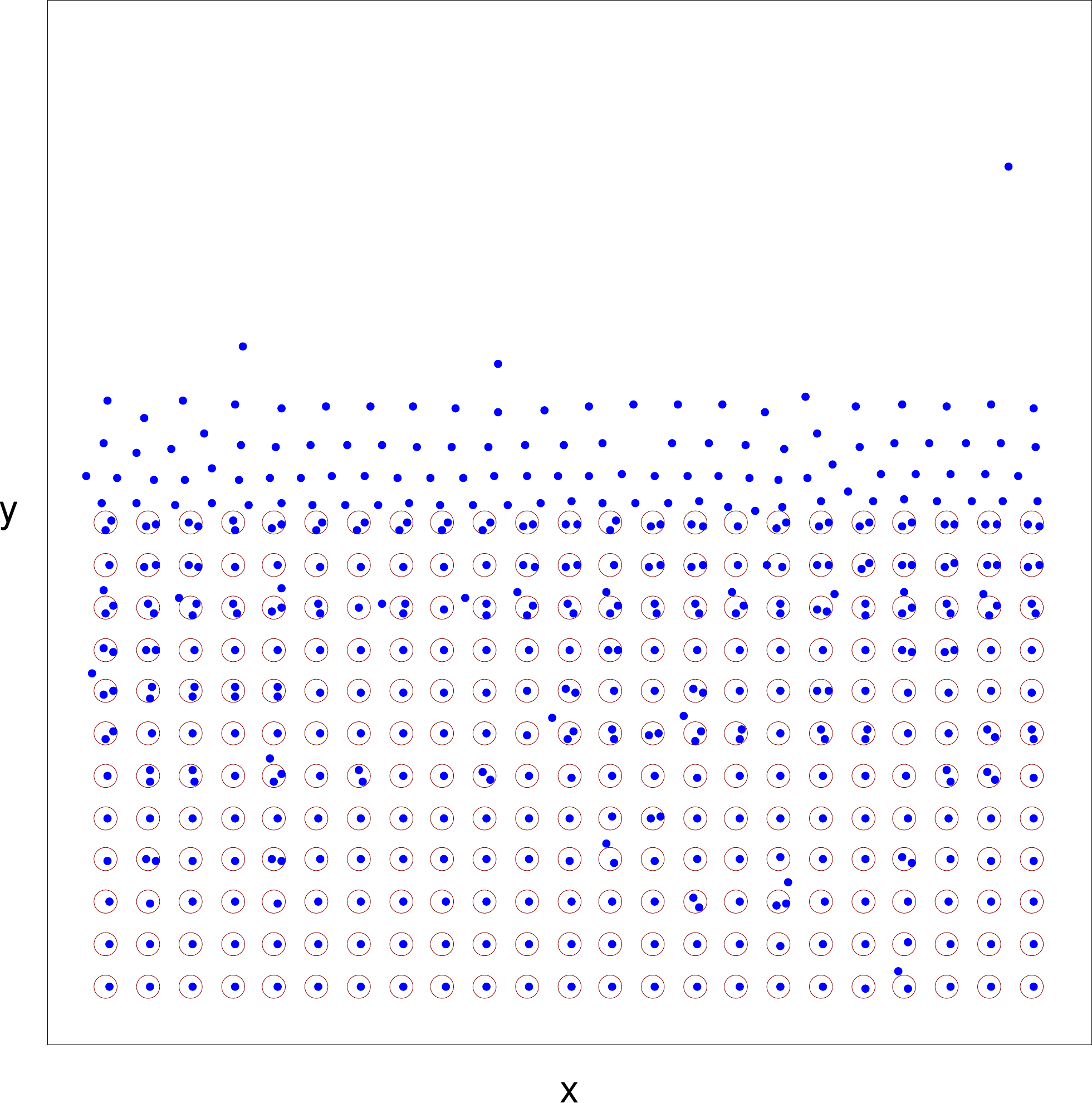}
\caption{Skyrmion locations (blue dots) and pinning site locations (open circles)
  for the system in Fig.~\ref{fig:25}(a) with $x$ direction driving at 
  $\alpha_m/\alpha_d=1.0$ and $F_p=4.5$ for $F_D=1.0$,
  just after the drop down in $\langle V_{||}\rangle$,
  showing multiple skyrmion  trapping by individual pinning sites.
}
\label{fig:26}
\end{figure}

As $F_{p}$ increases, we find wider windows of $F_D$ in which
$\theta_{sk}$ is small and gradually increasing.
When $F_{p} > 3.0$, we
start to observe the trapping of multiple skyrmions in individual pinning sites,
which causes the appearance of
intervals of $F_{D}$ in which $\langle V_{||}\rangle$ {\it decreases} with
increasing $F_D$.
This phenomenon is known as
negative differential conductivity
since $d\langle V_{||}\rangle/dF_{D} < 0$,
as illustrated
in Fig.~\ref{fig:25}(a) where we plot $\langle V_{||}\rangle$ and $\langle V_{\perp}\rangle$
versus $F_D$ for a system with $F_{p} = 4.5$ and $\alpha_{m}/\alpha_{d} = 1.0$.
Figure~\ref{fig:26} shows the skyrmion positions
in this sample at $F_D=1.0$,
just after the drop in $\langle V_{||}\rangle$,
where multiple skyrmions can be trapped at individual pinning sites
and where there is an increase in the skyrmion density along the boundary
of the pinned region.
In earlier work where the number of skyrmions
was much smaller than the number of pinning sites,
we also observed negative mobility that arises
when the skyrmions in the unpinned region are
forced into the pinned region and become immobile \cite{Reichhardt19c}.
A similar phenomenon can occur at higher skyrmion densities
when multiple skyrmions are trapped by each pinning site.
We also find signatures of multiple regimes in
$\theta_{sk}$,
as shown in Fig.~\ref{fig:25}(b).
The $\langle V_{||}\rangle$, $\langle V_{\perp}\rangle$, and $\theta_{sk}$
versus $F_D$ curves for
a system with $\alpha_m/\alpha_d=4.0$ appear in
Fig.~\ref{fig:25}(c,d),
highlighting the persistence of the
negative differential mobility effect
even for large intrinsic skyrmion Hall angles.  

\section{Driving Perpendicular to the Pinning Strip}

\begin{figure}
\includegraphics[width=3.5in]{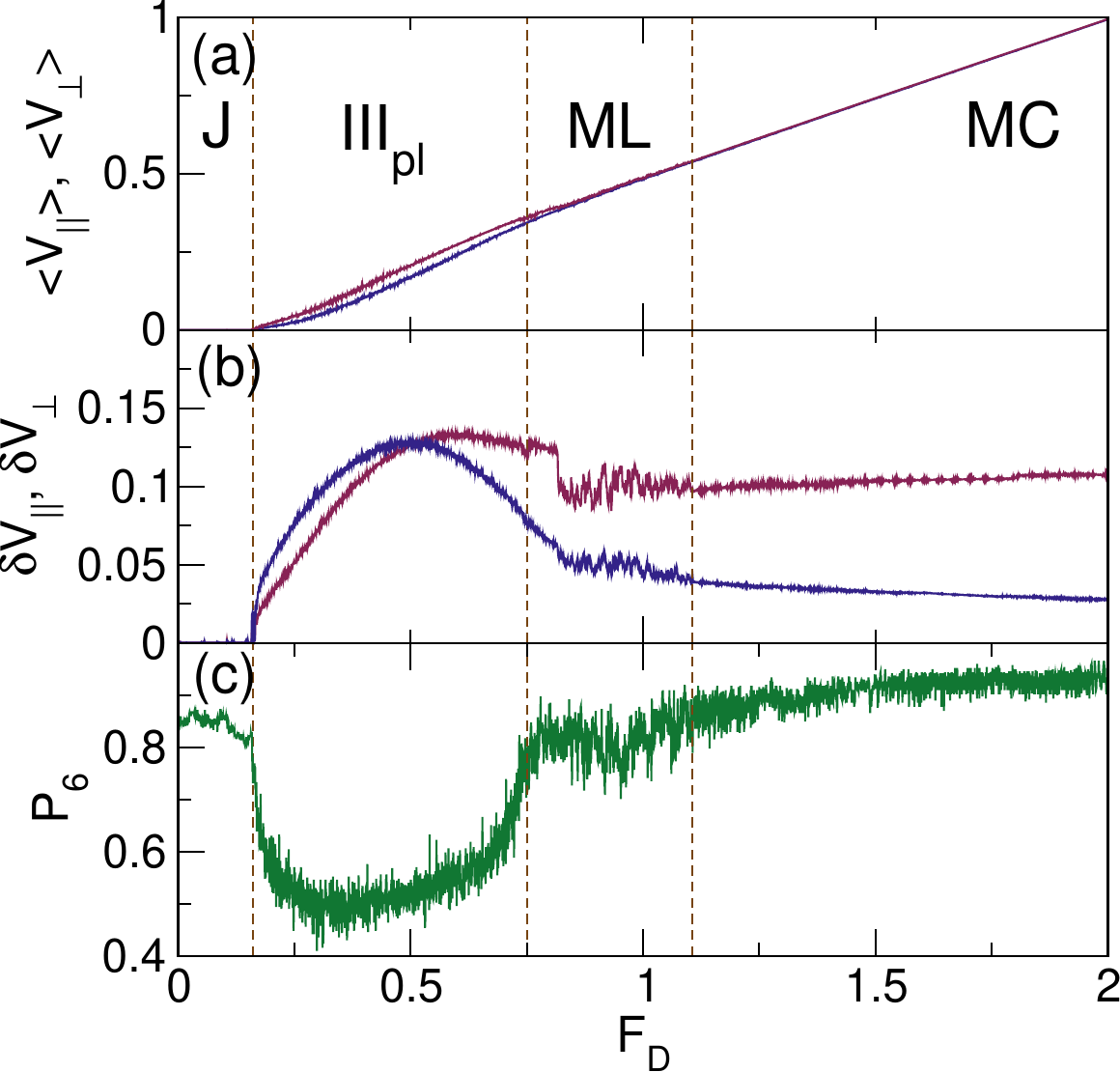}
\caption{Driving in the $y$ direction, perpendicular to the pinning strip,
  for the system in Fig.~\ref{fig:2} with
  $F_{p} = 0.75$ and $\alpha_{m}/\alpha_{d} = 1.0$.
  The vertical dashed lines denote
  transitions among a jammed phase J,
  a moving plastic flow phase III$_{pl}$, a ML, and a MC$_{sq}$ phase.
  (a) $\langle V_{||}\rangle$ (blue) and $\langle V_{\perp}\rangle$ (red) vs $F_{D}$.
  (b) $\delta V_{||}$ (blue) and $\delta V_{\perp}$ (red) versus $F_{D}$. 
  (c) $P_{6}$ vs $F_{D}$.
  Velocities are always measured parallel or perpendicular to the direction of
  the pinning strip.
}
\label{fig:27}
\end{figure}

\begin{figure}
\includegraphics[width=3.5in]{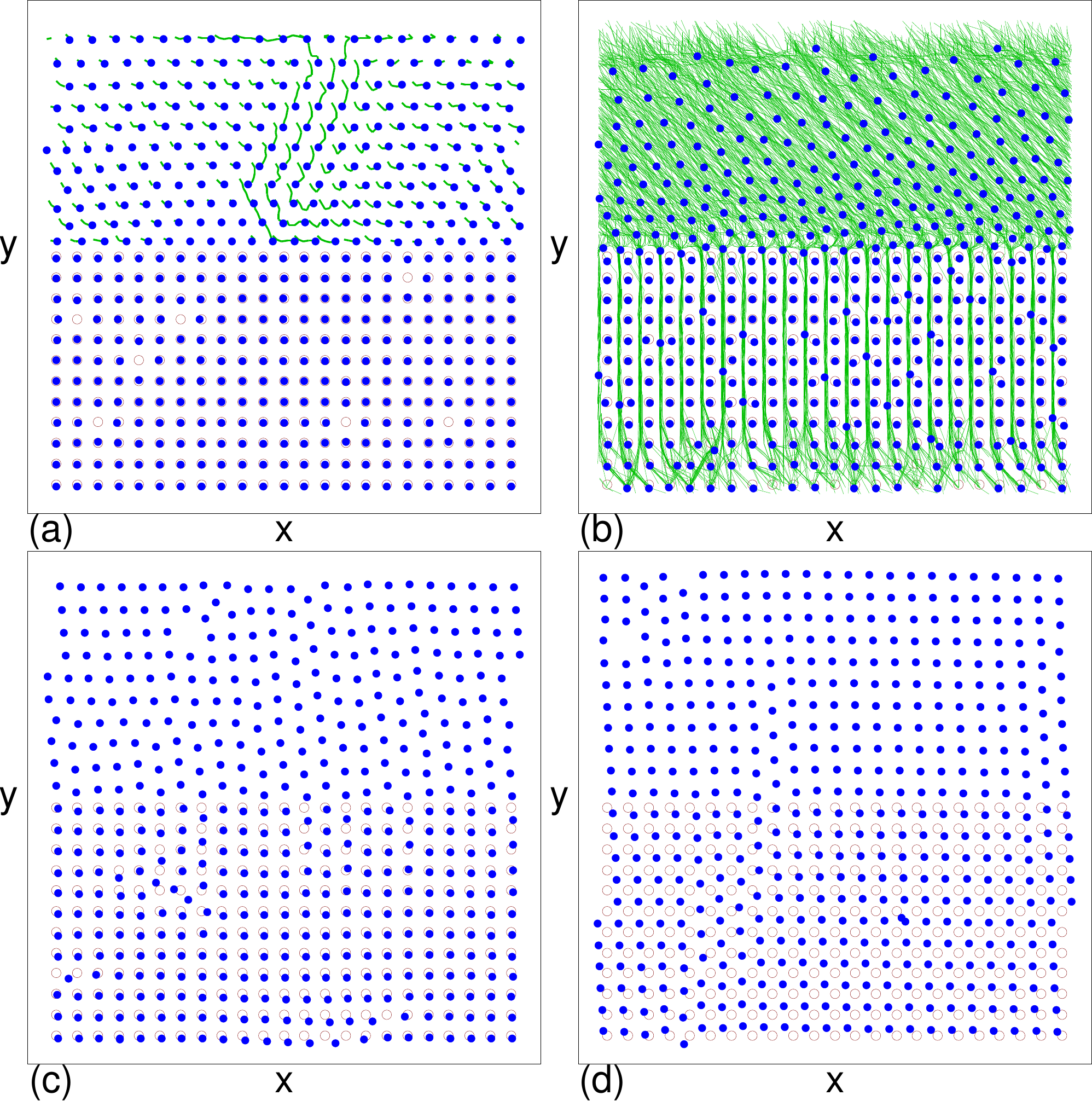}
\caption{Skyrmion locations (blue dots), pinning site locations (open circles),
  and skyrmion trajectories (green lines) during a fixed time interval
  for the system in Fig.~\ref{fig:27} with $y$ direction driving,
  $F_p=0.75$, and $\alpha_m/\alpha_d=1.0$.
  (a) Transient rearrangements in the jammed phase J at $F_{D} = 0.06$.
  (b) Phase III$_{pl}$ at $F_{D} = 0.25$, where the skyrmions can cross the
  entire system.
  (c) The ML phase at $F_D=1.0$,
  showing the skyrmion locations only.
  (d) The moving square lattice MC$_{sq}$ phase
  at $F_D=1.5$ showing the skyrmion locations only.  
 }
\label{fig:28}
\end{figure}

We next consider the effect of applying a drive along the $y$ direction,
perpendicular to the pinning strip. 
In an overdamped system, such a drive 
would push the particles into the pinned
region,
while for skyrmions, the Magnus force will also generate motion of the particles
parallel to the pin-free channel.
In general, we observe a jamming behavior
in which the skyrmions are pushed into the pinned region
but there is no steady state motion in either direction so that
$\langle V_{||}\rangle = \langle V_{\perp}\rangle=0$.
In Fig.~\ref{fig:27}(a) we plot $\langle V_{\perp}\rangle$  and $\langle V_{||}\rangle$
versus $F_D$ for a sample with $F_p=0.75$ and $\alpha_m/\alpha_d=1.0$
with $y$ direction driving.  As always, the parallel velocity is measured in the $x$
direction, parallel to the orientation of the pinning strip,
so here $\langle V_{\perp}\rangle$ shows motion in the direction of the applied drive.
We find a jammed phase J with $\langle V_{||}\rangle=\langle V_{\perp}\rangle=0$
for $F_{D} > 0.16$.
The jammed phase J is distinct from the pinned phase P described
earlier.
In phase P,
pinning results from
the finite shear modulus of the skyrmion lattice,
while in phase J, pinning arises due to the finite
compression modulus of the skyrmion lattice.
Since the compression modulus is much larger than the shear modulus,
the jammed phase appears over a much larger range of external drives
compared to the pinned phase.
Within the jammed phase,
temporary rearrangements or avalanches occur under increasing $F_D$
as the skyrmions adjust their positions to accommodate the drive,
as shown in Fig.~\ref{fig:28}(a) at $F_{D} = 0.06$. 
In Fig~\ref{fig:27}(b) we plot
$\delta V_{||}$ and $\delta V_{\perp}$ versus $F_D$.
Both quantities become finite in phase III$_{pl}$ when the skyrmions are
first able to travel all the way across the pinned region, as illustrated
in Fig.~\ref{fig:28}(b) at $F_{D} = 0.25$.
In Fig.~\ref{fig:27}(c), the $P_6$ versus $F_D$ curve has a drop
across the J-III$_{pl}$ transition.
As $F_{D}$ is further increased, the flow becomes more disordered.
For $0.75 < F_{D} < 1.1$, all the skyrmions are moving and the system forms a ML phase 
as shown in Fig.~\ref{fig:28}(c) at $F_{D} = 1.0$.
There is an increase in $P_6$ up to $P_6=0.8$
in the ML phase,
and we find strong
fluctuations in the velocity deviations $\delta V_{||}$ and $\delta V_{\perp}$
for $0.75 < F_{D} < 1.1$.
Above $F_D=1.1$,
the system enters a moving square crystal phase MC$_{sq}$,
illustrated in Fig.~\ref{fig:28}(d). The ML-MC$_{sq}$ transition
is visible as a shift in $P_6$ and a reduction of the fluctuations
in $\delta V_{||}$ and $\delta V_{\perp}$.

\begin{figure}
\includegraphics[width=3.5in]{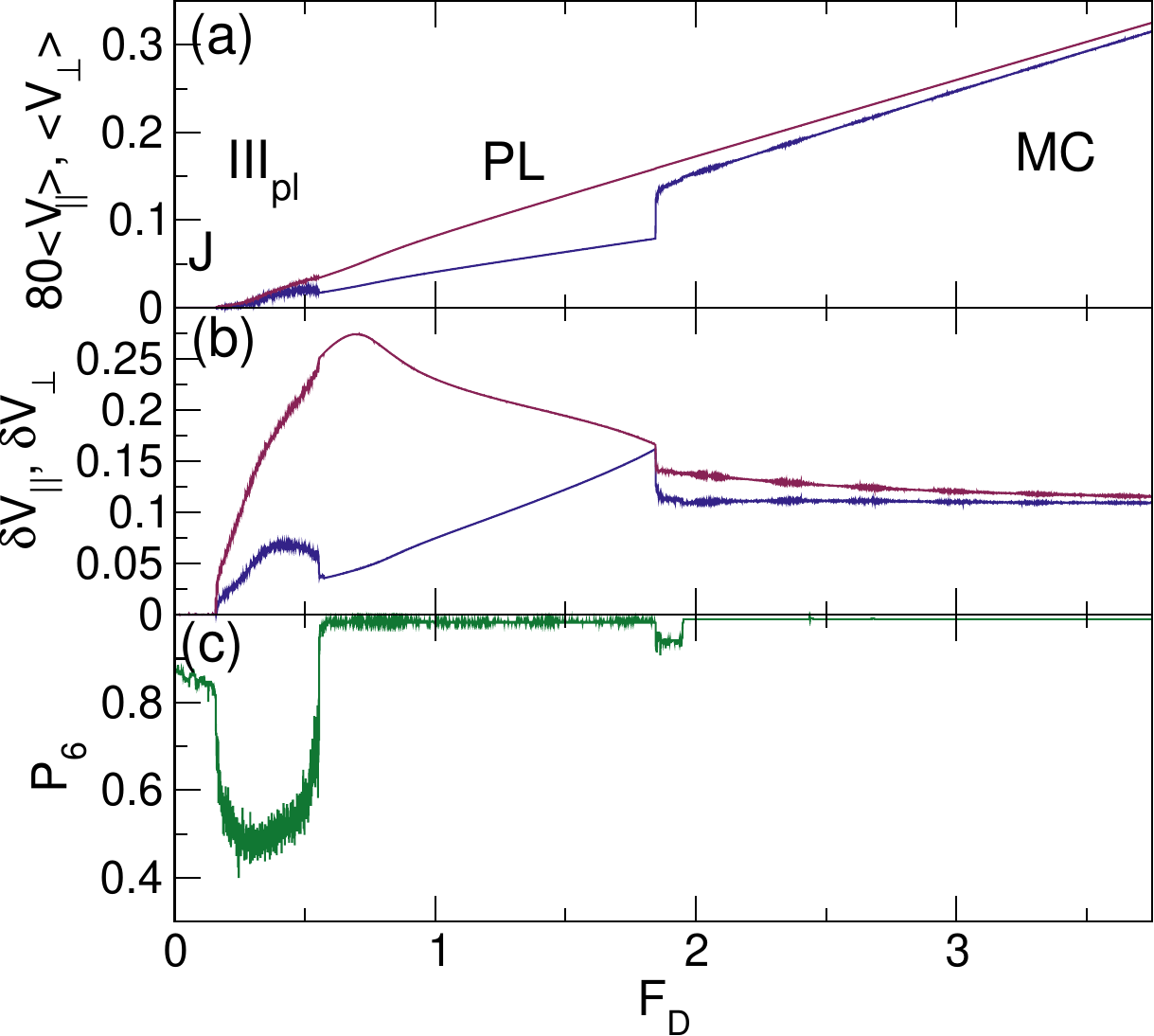}
\caption{ (a) $80\langle V_{||}\rangle$ (blue) and $\langle V_{\perp}\rangle$ (red)
  vs $F_{D}$ for the system
  with $y$ direction driving at $\alpha_{m}/\alpha_{d} = 0.0125$
  and $F_p=0.75$. 
  For clarity, we have divided $\langle V_{||}\rangle$ by the
  value of $\alpha_m/\alpha_d$ in order to normalize
  $\langle V_{||}\rangle$ to the value of
  $\langle V_{\perp}\rangle$ at high drives.
  (b) The corresponding $\delta V_{||}$ (blue) and $\delta V_\perp$ (red) vs $F_{D}$.
  (c) The corresponding $P_{6}$ vs $F_{D}$.
  We observe
  a jammed phase (J), plastic flow phase (III$_{pl}$),
  partially locked phase (PL),  and a moving crystal phase (MC).  
}
\label{fig:29}
\end{figure}

\begin{figure}
\includegraphics[width=3.5in]{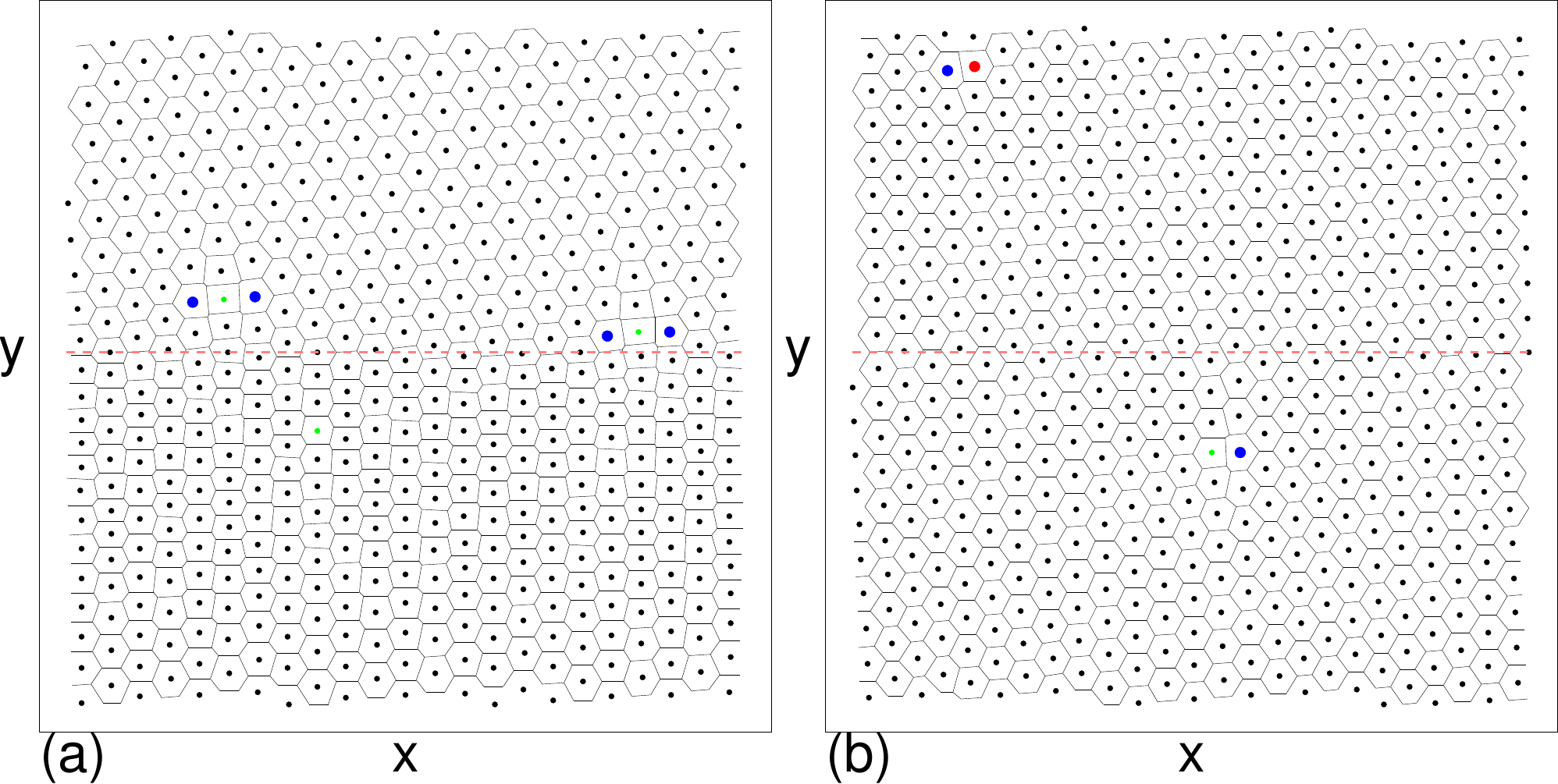}
\caption{ The Voronoi construction for the
  skyrmion locations for the system in
  Fig.~\ref{fig:29} with $y$ direction
  driving,
  $F_p=0.75$, and $\alpha_m/\alpha_d=0.0125$.
  The red dashed line indicates the edge of the pinned region.
  (a) The PL phase at $F_{D} = 1.0$, 
  where there is change in the orientation of the crystal across the boundaries separating
  the unpinned and pinned regions.
(b) The MC phase at $F_{D} = 2.5$.  
}
\label{fig:30}
\end{figure}

We find different regimes of behavior
for driving in the perpendicular direction
depending on whether $\alpha_{m}/\alpha_{d}$
is small or large.
In Fig.~\ref{fig:29}(a) we plot
$\langle V_{||}\rangle$ and $\langle V_{\perp}\rangle$ for a
system with $\alpha_{m}/\alpha_{d} = 0.0125$,
where we have normalized
$\langle V_{||}\rangle$ by dividing it by $\alpha_m/\alpha_d$.
Figures~\ref{fig:29}(b,c) show the corresponding
$\delta V_{||}$, $\delta V_{\perp}$, and $P_{6}$ versus $F_{D}$ curves. In this
case, we find a  jammed phase J and a plastic flow phase III$_{pl}$ in addition
to 
a new phase which
we call a partially locked (PL) state.
Here,
the motion of the skyrmions within the pinned region is completely
locked in the $y$-direction, parallel to the drive, but the skyrmions in the unpinned
region move at an angle to the drive.
The III$_{pl}$-PL phase transition is associated with
an increase in $P_{6}$ and a drop in $\langle V_{||}\rangle$.
In phase III$_{pl}$, skyrmions are moving in both the $x$ and $y$ directions;
however, in the PL phase, skyrmions in the pinned region move only in the $y$
direction, causing the value of $\langle V_{||}\rangle$ to drop.
In Fig.~\ref{fig:30}(a) we show the Voronoi constriction for the
skyrmion locations in the PL phase at $F_{D} = 1.0$.
Within the pinned region, the skyrmion lattice
is aligned with the $y$-direction,
while in the unpinned region, it is aligned at an angle to the $y$
direction.
Near $F_{D} = 2.0$ in Fig.~\ref{fig:29},
there is a transition to a MC phase in which
the skyrmion motion is no longer locked
to the $y$ direction,
as indicated by a jump up in $\langle V_{||}\rangle$ and
a cusp in $\delta V_{||}$ and $\delta V_{\perp}$.
There is also a small dip in $P_{6}$ near the PL-MC transition.
Figure~\ref{fig:30}(b) shows the Voronoi construction of the
skyrmion locations
in the MC phase at $F_{D} = 2.5$,
where the entire skyrmion lattice is aligned in the same direction. 
The PL phase is
produced by the locking of the skyrmion motion along the symmetry direction of
the pinning lattice,
but when $\alpha_{m}/\alpha_{d}$ is large enough, the partially locked phase is 
replaced by the moving crystal phase.

\begin{figure}
\includegraphics[width=3.5in]{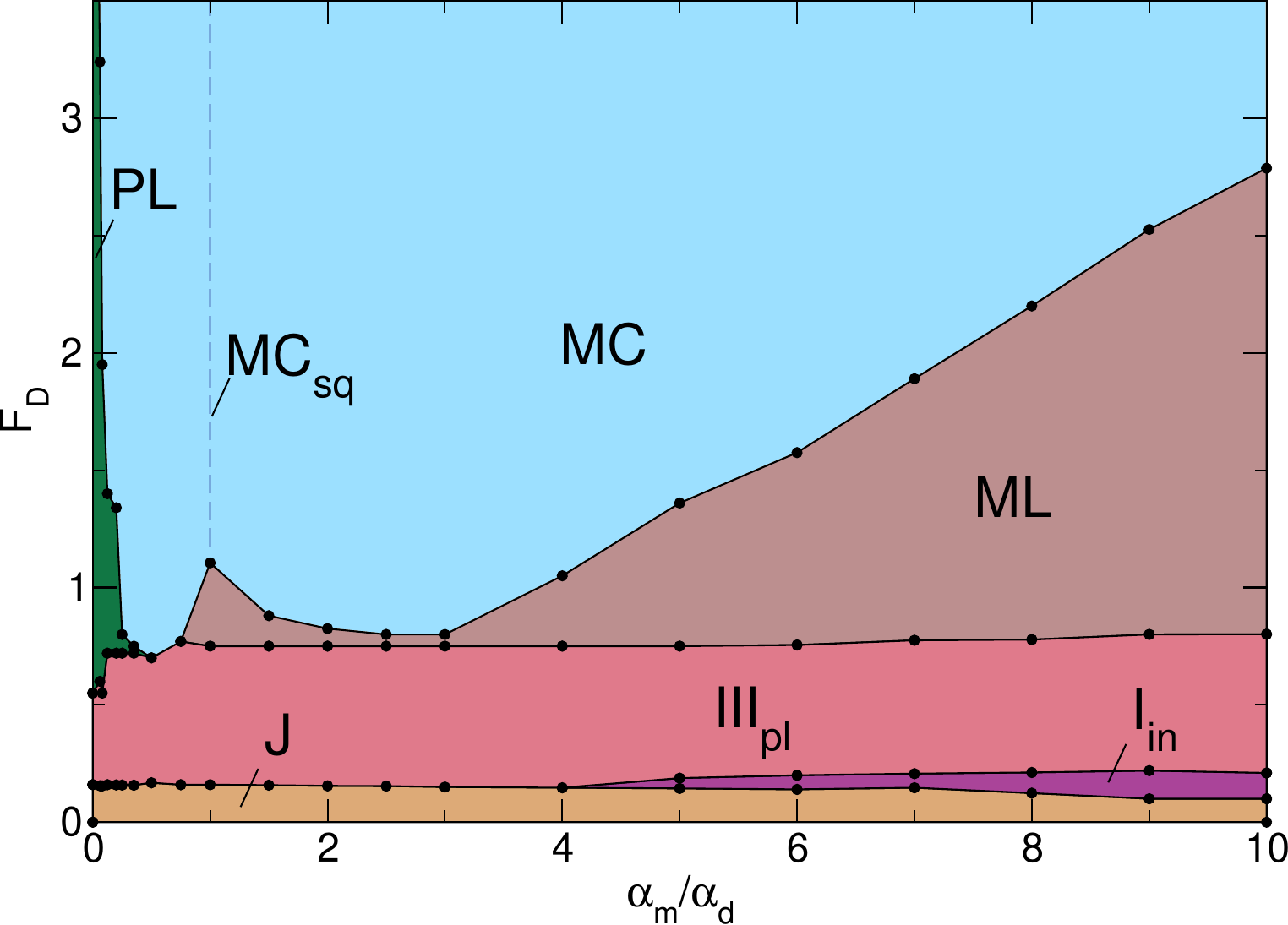}
\caption{Dynamic phase diagram as a function of
  $F_{D}$ vs  $\alpha_{m}/\alpha_{d}$ for driving in the $y$ direction
  for a system with $F_{p} = 0.75$.
  Partially locked phase (PL), dark green;
  jammed phase (J), tan;
  plastic flow (III$_{pl}$), pink;
  intermittent (I$_{in}$), violet;
  moving liquid (ML),  brown;
  moving crystal (MC), light blue.
  At $\alpha_{m}/\alpha_{d} = 1.0$ there
is a moving square lattice (MC$_{sq}$), dark blue dashed line.    
}
\label{fig:31}
\end{figure}

In Fig.~\ref{fig:31} we construct a dynamic phase diagram
as a function of $F_{D}$ versus $\alpha_{m}/\alpha_{d}$ for
driving in the $y$ direction.
The PL phase
appears when $\alpha_{m}/\alpha_{d} <  0.4$
and diverges in width as $\alpha_{m}/\alpha_{d}$ goes to zero.
At low drives, the jammed phase J occurs for
all $\alpha_m/\alpha_d$,
and is followed at higher drives by
an intermittent avalanche phase I$_{in}$
when $\alpha_{m}/\alpha_{d} \geq 5.0$.
The I$_{in}$ phase
is similar to that found
for driving in the $x$
direction but its onset is at
somewhat lower values of 
$\alpha_{m}/\alpha_{d}$.
We find the plastic phase III$_{pl}$ for
all values of $\alpha_m/\alpha_d$,
and the ML phase increases in extent 
with increasing Magnus force,
similar to the behavior of the ML phase in the system with $x$ direction
driving.
At $\alpha_{m}/\alpha_{d} = 1.0$, there is a transition from
the ML
into a moving square lattice
rather than to a moving crystal phase.
It is possible that for $\alpha_{m}/\alpha_{d}$ near $1.0$,
there could be a regime in which
the moving square lattice transitions into the moving crystal phase,
with the value of $F_D$ at which the transition occurs diverging at
$\alpha_{m}/\alpha_{d} = 1.0$. 
For fixed $\alpha_{m}/\alpha_{d}$ and increasing $F_{p}$,
we observe a similar evolution of the phases as
in the system with $x$ direction driving,
including a transition to elastic depinning
for small $F_p$.
 
\section{Discussion}
Our results should be general for skyrmion systems 
with some form of inhomogeneous pinning.
Although we specifically focus on the case of a square lattice of pinning sites,
we expect similar results to appear
if the pinned region contains randomly placed pinning sites
or a triangular pinning lattice.
One
exception
to this is that
the strong guidance
effects that occur for skyrmion Hall angles of
$\theta^{\rm int}_{sk} = 45^\circ$ would not appear for most other pinning
geometries since these are specific to the square pinning lattice.
In this work we did not consider hysteretic effects, but it is likely 
that many of the phases would show hysteresis while others would not.
There is now evidence for a variety of skyrmion systems
that exhibit thermal effects or diffusion, so it would be
interesting to study
how temperature could affect the results, such as by inducing creep.
It may also be possible to measure the evolution of the shear
modulus with temperature.
For instance,
at low drives there is a pinned phase in which the skyrmions
trapped at the pinning sites hold back the skyrmions in the pin-free region,
and this can only occur if the skyrmion lattice has a finite shear modulus.
As the temperature increases, the shear modulus would be reduced
before vanishing at the melting transition, which would destroy the indirect
pinning of the skyrmions in the pin-free region, causing them to flow.
Similar effects have been 
studied in the context of superconducting vortices
driven through weak pinning channels.
Although our work involves skyrmions, many of these results could be
relevant
to other systems even in the absence of a Magnus force.
Some examples include
vortices or colloids in a system with a combination of strong and weak pinning
where the direction of the drive is not fixed but gradually changes,
which could mimic the effect of the changing skyrmion Hall angle.

\section{Summary}
We have numerically examined the dynamics of skyrmions in systems with inhomogeneous pinning. We focus on a sample containing
a strip of square pinning coexisting with a region of no pinning.
When the external
drive is parallel to the pinning strip,
we find that initially, only the skyrmions in the unpinned region flow,
termed phase I motion,
and that at higher drives, a shear
banding phase II$_{sb}$ emerges in which
flow occurs in the pinned region with a velocity gradient.
Both of these phases have a skyrmion Hall angle of zero.
When the drive is strong enough,
the skyrmions enter the plastic motion or disordered phase III$_{pl}$ in which
the skyrmion Hall angle becomes finite.
At higher drives, there is a moving liquid (ML) phase followed by a transition into a
moving crystal (MC).
The skyrmion Hall angle increases most rapidly
with increasing drive in the plastic flow phase, and increases more slowly in the ML phase. 
For phases I, II$_{sb}$, and III$_{pl}$, there is an accumulation of skyrmions along the
edge of the pinned region
due to the Magnus force, which pushes the skyrmions in the unpinned region toward
the pinned region.
As the strength of the Magnus force increases,
there is a phase in which
the skyrmions enter the pinned region under avalanche-like transport,
creating a density gradient similar to the Bean state found in 
type-II superconductors.
There is also an intermittent or dynamically phase separated state
in which the skyrmion motion through the pinned region is confined to
localized rivers that change in size and location with increasing drive.
For weak Magnus forces, we find a smectic phase in which the skyrmion motion
is locked
in the direction of the drive over an extended range of driving forces,
followed by a
transition to a state in which the skyrmion Hall angle is finite.
For weak pinning, the
system exhibits an elastic pinning regime
with 
a shear jammed state at low drives followed
at higher drives by a moving crystal
phase in which each skyrmion keeps its same neighbors as it moves.
When the pinning is strong,
we observe negative differential conductivity when
skyrmions in the pin-free channel
are pushed into the pinned region and become immobile,
dropping the overall mobility of the system. 
When the external drive is perpendicular to the pinning strip,
we find a jammed phase at low drives in which
skyrmions in the unpinned region are unable to move into the pinned region.
As the drive increases,
there is a partially locked phase,
a moving liquid state, and a moving crystal phase.
We show how
the transitions between these different phases
produce signatures in
the skyrmion mobility, coordination numbers,
velocity deviations,
and global skyrmion lattice structure.
Beyond skyrmions,
our results could also be relevant for other systems in which Magnus forces
can arise,
such as vortices in superconductors or superfluids as well as in
chiral active matter systems.  

\acknowledgments
This work was supported by the US Department of Energy through
the Los Alamos National Laboratory.  Los Alamos National Laboratory is
operated by Triad National Security, LLC, for the National Nuclear Security
Administration of the U. S. Department of Energy (Contract No. 892333218NCA000001).

\bibliography{mybib}
\end{document}